\documentclass[useAMS,usenatbib,fleqn]{mnras}
\usepackage{amsmath}
\usepackage{multirow}
\usepackage{graphicx}
\usepackage{color,soul}
\usepackage{epsfig}
\usepackage{amssymb,amsmath}
\usepackage{aas_macros}
\usepackage{fixltx2e}
\title[MBHs and CIB Fluctuations]{
The Clustering of Undetected High-redshift Black Holes and Their Signatures in Cosmic Backgrounds
}

\author[Ricarte et al.]{Angelo Ricarte$^1$, Fabio Pacucci$^{1,2,3}$, Nico Cappelluti$^{4,5}$, Priyamvada Natarajan$^{1,2}$ \\
$^1$ Department of Astronomy, Yale University, 52 Hillhouse Avenue, New Haven, CT 06511 \\
$^2$ Department of Physics, Yale University, P.O. Box 208121, New Haven, CT 06520 \\
$^3$ Kapteyn Astronomical Institute, University of Groningen, Landleven 12, 9747 AD Groningen \\
$^4$ Department of Physics, University of Miami, Coral Gables, Florida 33124 \\
$^5$ INAF-Osservatorio di Astrofisica e Scienza dello Spazio di Bologna, via Gobetti 93/3, I-40129 Bologna, Italy
}

\date{\today}

\begin{document}
\pagerange{\pageref{firstpage}--\pageref{lastpage}} \pubyear{2019}
\maketitle

\begin{abstract}
There exist hitherto unexplained fluctuations in the Cosmic Infrared Background (CIB) on arcminute scales and larger.  These have been shown to cross-correlate with the Cosmic X-ray Background (CXB), leading several authors to attribute the excess to a high-redshift growing black hole population. In order to investigate potential sources that could explain this excess, in this paper, we develop a new framework to compute the power spectrum of undetected sources that do not have constant flux as a function of halo mass. In this formulation, we combine a semi-analytic model for black hole growth and their simulated spectra from hydrodynamical simulations. Revisiting the possible contribution of a high-redshift black hole population, we find that too much black hole growth is required at early epochs for $z>6$ accretion to explain these fluctuations.  Examining a population of accreting black holes at more moderate redshifts, $z\sim 2-3$, we find that such models produce a poor fit to the observed fluctuations while simultaneously overproducing the local black hole mass density. Additionally, we rule out the hypothesis of a missing Galactic foreground of warm dust that produces coherent fluctuations in the X-ray via reflection of Galactic X-ray binary emission.  Although we firmly rule out accreting massive black holes as the source of these missing fluctuations, additional studies will be required to determine their origin.
\end{abstract}


\section{Introduction}\label{sec:intro}

%Move this spacing around in order to work around the citation-split compilation bug.
\vspace{0pt}

The Cosmic Infrared Background (CIB) and Cosmic X-ray Background (CXB) contain the integrated emission of all photons in the universe from starlight and accretion powering active galactic nuclei (AGN) \citep{Salvaterra&Ferrara2003,Cooray+2004,Kashlinsky+2004}.  Both the CIB and the CXB are composed of a resolved component detectable in deeper surveys, and an unresolved component arising from sources that are not individually detected.  The unresolved component of the CIB is believed to originate primarily from undetected low luminosity galaxies \citep{Kashlinsky+2012,Cooray+2012,Yue+2013a}, while the unresolved CXB likely originates from faint populations of as yet undetected AGN and star-forming galaxies \citep{Gilli+2007}.  This is inferred by evaluating the putative contribution from extrapolations of the following known source populations: star-forming galaxies, AGN, and the hot diffuse gas in groups and clusters. 

In the near-infrared, the absolute intensity of the CIB is difficult to infer due to the dominant contribution of zodiacal light and the uncertain contribution of Galactic cirrus \citep{Leinert+1998}.  While zodiacal light dominates the total intensity, it is fortunately thought to be very structurally smooth \citep{Kelsall+1998}.  Therefore, studying CIB {\it fluctuations} provides an alternative approach to determine its origin \citep{Kashlinsky+1996}.  Fluctuation analyses of the measured CIB---its power spectrum or two-point correlation function---have revealed excess power on arcminute to degree scales on the sky that remain unexplained even with the most optimistic extrapolations of faint galaxy populations \citep{Helgason+2012,Yue+2013a}, pointing to an entirely new missing contribution to the inventory.  This excess has now been seen in both the 3.6 and 4.5 micron bands by two different infrared telescopes \citep{Kashlinsky+2005,Kashlinsky+2012,Matsumoto+2011}, is detected in multiple independent fields, and has persisted through years of improved measurements \citep[see][for a review]{Kashlinsky+2018}.  

Analyses of the missing fluctuations' electromagnetic spectrum offer some additional clues to their origin.  Recently, these excess CIB fluctuations have been shown to cross-correlate with the fluctuations in the CXB at the $5\sigma$ level \citep{Cappelluti+2013,Mitchell-Wynne+2016,Cappelluti+2017b,Li+2018}.  This has prompted some authors to attribute this excess to accreting massive black holes (MBHs), by which we refer to the black holes at the centres of galaxies that power AGN.  A later cross-correlation analysis performed by \citet{Li+2018} found that the spectral energy distribution (SED) of the fluctuations is consistent with unobscured black hole accretion.  On the other hand, broad-band SEDs of these fluctuations in the near infrared (NIR) are consistent with a Rayleigh-Jeans spectrum \citep{Matsumoto+2011}.  Hence, other authors have attributed the excess to stars in these extra-galactic haloes.  Currently, there is consensus that this excess power is real and astrophysical in nature, but its origin remains unsettled and controversial. 

\citet{Helgason+2014} perform the most comprehensive study to date of the known source populations contributing to both the CIB and CXB.  They determine that while extrapolations of known galaxy and AGN populations can explain the coherence of the backgrounds on scales $<1'$, only 3 per cent of the measured cross-power between the CIB and CXB on larger angular scales can be obtained from known populations at redshifts $z \leq 5$.  Based on their luminosity functions, known populations of AGN are expected to contribute negligibly to the CIB auto-power fluctuations.  Theoretical estimates of galaxy populations at yet higher redshifts also fall short of explaining the excess by an order-of-magnitude \citep{Helgason+2012,Yue+2013a}.  

The two predominant hypotheses for the explaining the proposed excess with new populations are intra-halo light (IHL) \citep{Cooray+2012} and $z>10$ Compton-thick accreting MBHs \citep{Yue+2013b}.  In the IHL interpretation, there exists uncharacterised starlight beyond the radius at which galaxies are masked.  This picture is tantalizing due to the Rayleigh-Jeans-like spectrum of this component in the NIR.  On the other hand, the background fluctuations do not appear to be sensitive to the size of the mask \citep{Arendt+2010}.  More importantly, this model does not provide a mechanism to generate a cross-correlation with the CXB.  While X-ray emitting halo gas could hypothetically cross-correlate with the IHL, it would not create enough hard X-rays to be consistent with the amplitude of the fluctuations \citep{Li+2018}.  If it exists, the IHL component remains to be characterised by either cosmological simulations theoretically or low-surface brightness observations.

Alternatively, an early population of $z > 6$ Compton-thick AGN could naturally explain the cross-correlation with the X-ray \citep{Yue+2013b,Cappelluti+2013,Cappelluti+2017b}.  The Compton-thick nature of these objects is required to prevent them from exceeding the X-ray background and causing reionisation prematurely. However, as we show, in order to produce sufficient integrated intensity to match the observed excess, such a population of accreting MBHs must accrete an amount of matter even greater than the $z=0$ mass density of MBHs \citep[e.g.,][]{Yue+2013b} at these early epochs.  This is not expected or predicted in any models that track early MBH growth, which in fact generally struggle to grow MBHs rapidly enough to explain $z\sim 6$ quasars without special conditions such as massive seeds, super-Eddington accretion rates, and galactic inflows that are not curtailed by feedback \citep[e.g.,][]{Haiman&Loeb2001,Volonteri+2015,Pacucci+2017c,Woods+2018}.  Another hypothesized solution also utilizes early black holes, but indirectly.  If dark matter is composed of primordial black holes \citep{Bird+2016}, extra Poissonian fluctuations in the early universe would lead to isocurvature density fluctuations that create additional low-mass haloes that could contribute to the background \citep{Kashlinsky+2016}.

Previous work thus far has not employed self-consistent models of MBH assembly tracked over cosmic time that are constrained by the wealth of currently available multi-wavelength data.  To decisively settle whether a population of early MBHs could indeed explain the observed excess, we adopt a new comprehensive, self-consistent, and empirically motivated approach that enables the calculation of correlation functions. The standard approach at present, referred to as a semi-analytic model (SAM), has been very successfully used to track the co-evolution of dark matter haloes, galaxies and their MBHs, standing as an established and robust method to interpret galaxy and AGN surveys \citep[e.g.,][]{Kauffmann+1993,Volonteri+2003,Somerville+2008,Benson+2012}.  Using our comprehensive modelling, we find that a realistic $z>6$ MBH population falls many orders of magnitude short in flux to explain the observed excess.  We then explore alternative models: first, a population of buried $z\sim 2-3$ AGN, and then a simple power-law which may be due to a missing local component.  We rule out an AGN explanation to the missing fluctuations, but also argue that no other competing physical model can explain all of the observations.

In \S\ref{sec:data}, we describe the data that we use, which we adopt without reanalysis.  In \S\ref{sec:methodology}, we establish our theoretical framework, which consists of a semi-analytic model for MBH growth \S\ref{sec:sam}, simulated spectra from hydrodynamical simulations \S\ref{sec:gems}, and a new formalism to describe the clustering of undetected sources \S\ref{sec:clustering}.  The results of our calculations are shown in \S\ref{sec:results_power_spectra}, where we show that $z>6$ MBH accretion cannot be responsible for the excess.  In \S\ref{sec:discussion}, we explore alternate explanations: the more moderate-redshift, buried AGN as well as a new Galactic foreground.  Our work is summarized in \S\ref{sec:conclusion}.

\section{Data}\label{sec:data}

We utilize the most recent and carefully estimated power spectrum measurements of \citet{Cappelluti+2017b} without performing any reanalysis.  These include measurements of the cross-power spectra of the unresolved Spitzer-IRAC cosmic infrared background and 0.5-2 Chandra-ACIS CXB fluctuations. These include data from the Chandra Deep Field South, Hubble Deep Field North, Extended Groth Strip/AEGIS field, and UDS/SXDF surveys, where a $>5-6 \sigma$ cross-power signal is detected on angular scales $>20$'' between [0.5-2] keV and the 3.6 and 4.5 $\mu$m bands, as well as 3.6+4.5 $\mu$m combined.  No significant correlation was found with harder X-ray bands.  Moreover, they find an excess of about an order of magnitude at $5\sigma$ confidence when compared to models for the contribution of the known unmasked source population at $z<7$ from \citet{Helgason+2014}. \citet{Mitchell-Wynne+2016} detected a similar signal in the Chandra Deep Field South.  Recently, \citet{Li+2018} extended and confirmed these results and detected the signal on even larger scales - up to scales of $\sim$1000''-1500'' using the Spitzer-SPLASH survey and the Chandra-COSMOS legacy survey data. The signal even on these scales is still in excess with respect to that produced by foreground sources.  

IR maps were self-calibrated and Zodiacal light removed following \citet{Arendt+2010}. Resolved IR sources were subtracted down to AB magnitude $m\sim 25$ in both the 3.6 and 4.5 $\mu$m IRAC bands \citep{Kashlinsky+2012}.  0.5-2 keV, X-ray maps of each individual field were particle subtracted and masked from sources brighter than a few $10^{-17} \ \mathrm{erg} \; \mathrm{cm}^{-2} \; \mathrm{s}^{-1}$.  These maps were employed to measure the cross-power spectrum with Fourier analysis. Errors are estimated with a simple cosmic variance estimator \citep[see][]{Cappelluti+2013}. 

In addition to the cross-power signal, \citet{Li+2018} were able to determine that the X-ray spectrum of the CXB fluctuations is coherent with those in the CIB and found it can be fitted by either a power-law component with spectral index $\gamma = 2$ or with a $z\sim 7-15$ population of absorbed Compton-Thick AGN or with a direct collapse black hole (DCBH) template from \citet{Pacucci+2015}. 

\section{Methodology}\label{sec:methodology}

\begin{figure*}
    \centering
    \includegraphics[width=\textwidth]{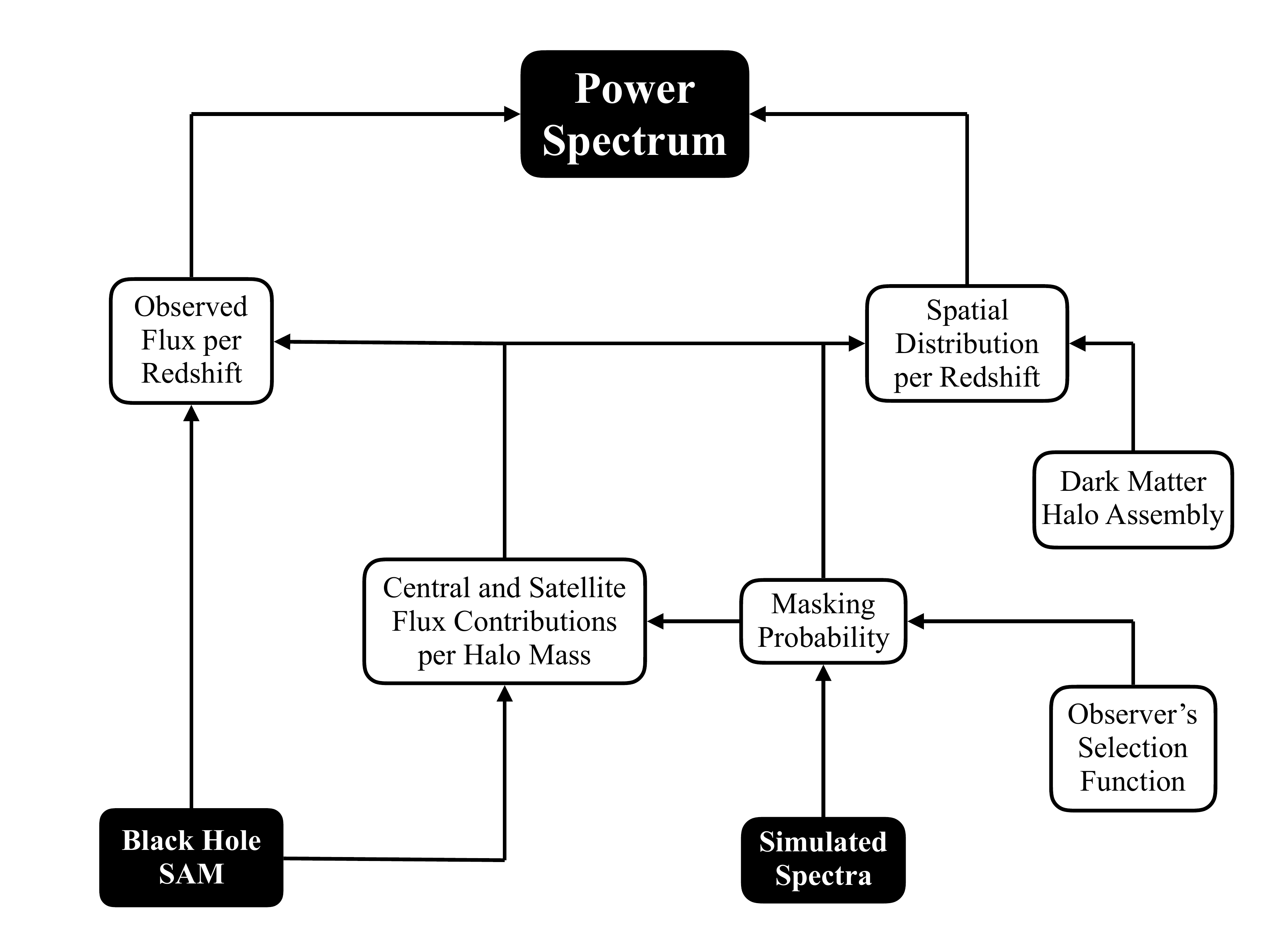}
    \caption{Schematic figure detailing our modelling methodology.  We combine a black hole semi-analytic model with synthetic spectral energy distributions in order to compute the angular power spectrum of the background. Since our MBHs are undetected, we model the distribution of flux rather than the distribution of sources, accounting for observational selection effects.
    \label{fig:computing_power_spectrum}}
\end{figure*}

We develop a comprehensive theoretical framework to predict the MBH contribution to the power spectra of the CIB and CXB.  We begin with a semi-analytic model for MBH growth from the seeding epoch to the present day \citep{Ricarte&Natarajan2018a,Ricarte&Natarajan2018b}.  For each individual MBH in the model, we compute its contribution to the CIB and CXB using hydrodynamical simulations to extract the multi-wavelength spectra of early MBHs.  We begin with the standard halo occupation distribution (HOD) methodology, and derive a new formalism to determine the power spectrum of sources that are not individually detected.  Our framework and approach are illustrated graphically in Fig. \ref{fig:computing_power_spectrum}.  Below, we explain each aspect of this modelling in more detail.  Note that in this work we do {\it not} consider a hypothetical primordial black hole population as in \citet{Kashlinsky+2016}.

\subsection{Semi-analytic Modelling}\label{sec:sam}

We begin with a semi-analytic model for the assembly of MBHs that starts at the seeding epoch.  We explore both the ``light'' and ``heavy'' seeding models of \citet{Ricarte&Natarajan2018a,Ricarte&Natarajan2018b}.  These models integrate the evolution of MBHs from $z=20$ to $z=0$ using simple prescriptions that relate the evolution of MBHs to their host galaxies.  They have been tested against the local $M_\bullet-\sigma$ relation \citep[relating MBH mass to host velocity dispersion][]{Kormendy&Ho2013,Saglia+2016,vandenBosch2016} and bolometric luminosity functions for $0<z<6$ \citep{Hopkins+2007}.  A hybrid approach is taken whereby empirical relations are used for the galaxy-halo connection, while analytic prescriptions are used to model the MBH-galaxy connection.  

First, dark matter merger trees are constructed for a statistically representative sample of the universe, using an extended Press-Schechter algorithm \citep{Press&Schechter1974} calibrated to the Millennium simulations \citep{Parkinson+2008}.  We first estimate the contribution of the undetected high redshift population of growing black holes in our the context of our model. Since we are interested only in the high-redshift universe, the parent nodes of these merger trees originate at $z=4$. We sample 35 halo masses between $10^{10-13.4} \ \mathrm{M}_\odot$, 100 times each in order to capture cosmic variance.  This wide range of halo masses includes even highly biased $5\sigma$ peaks which are implicated as the sources of the most distant quasars \citep{Fan+2003}.

With these trees in hand, we then integrate the assembly of MBHs from $z=20$ to $z=4$.  Black hole seeding is permitted in the redshift range $15<z<20$, when the pristine nature of gas promotes the formation of objects with large Jeans masses \citep{Bromm+2002,Agarwal&Khochfar2015}.  In light seed models, any halo which represents a 3.5$\sigma$ peak or higher is randomly assigned a 30-100 $\mathrm{M}_\odot$ seed \citep{Stacy+2016}.  In heavy seed models, direct collapse black hole seeds in the mass range $10^{4-6} \ \mathrm{M}_\odot$ form in haloes with low spins and virial temperatures appropriate for atomic-cooling haloes \citep{Lodato&Natarajan2006}.  A detailed description of our seeding model can be found in \citep{Ricarte&Natarajan2018b}.  The origin of seed black holes is currently an open question and observations suggest that more than one channel to produce early seeds is likely to operate in the universe \citep{Volonteri&Bellovary2012}. Therefore, investigating potential differences between light and heavy seed populations that might enable discrimination between them is warranted \citep{Natarajan2014,Pacucci+2018,Natarajan+2019}.  We note here that there are no numerical simulations currently available that can self-consistently track seed formation and growth over cosmologically relevant volumes and then compute measurable quantities like clustering statistics - therefore, we rely currently on semi-analytic modelling.

As in previous work with this fiducial model, major mergers trigger MBH growth until MBHs reach the the $M_\bullet-\sigma$ relation, where $\sigma$ represents the stellar velocity dispersion within the effective radius of the host galaxy.  In order to directly compare with the observational data, it is important that $\sigma$ {\it not} be the velocity dispersion of the dark matter halo, which evolves differently with redshift. We instead estimate $\sigma$ based on a combination of the stellar-to-halo mass relation \citep{Moster+2013} and the stellar mass-to-size relation \citep{Mosleh+2013,Huertas-Company+2013}.  In previous work, we had set the black hole growth rate to the Eddington rate during this period.  In this work, we instead draw Eddington ratios randomly from the observed distribution of observed broad-line quasars \citep{Kelly&Shen2013,Tucci&Volonteri2017}.  These distributions allow for some super-Eddington accretion at high-redshift, but again, only until MBHs reach the $M_\bullet-\sigma$ relation.  In Appendix \ref{sec:model_validation}, we show that these accretion prescriptions allow us to match luminosity functions at $0<z<6$ (for a separate set of low-redshift merger trees) as well as obtain reasonable values for the clustering of AGN at low-redshift that are in agreement with observational data.

Once these calculations are complete, we have a large population of haloes with known AGN bolometric luminosities. These bolometric luminosities are self-consistently related to MBH accretion rates via $L_\mathrm{bol} = \epsilon_r \dot{M}_\bullet c^2$, where $\epsilon_r$ is the radiative efficiency (set as usual to 0.1) and $c$ is the speed of light.  To proceed, we now need to determine what fractions of these bolometric luminosities are emitted in the relevant bands, and additionally we develop a formalism for calculating the clustering of undetected sources.

\subsection{From MBH Growth Rates to Fluxes} \label{sec:gems}

\subsubsection{The Spectra of High-z Black Hole Seeds}

The contribution of MBHs to the CXB/CIB depends on the underlying spectral template assumed for the output emission of these accreting sources.  Rather than simply treating these bolometric corrections as free parameters, we employ the radiation-hydrodynamic code GEMS \citep[Growth of Earth Massive Seeds][]{Pacucci+2015,Pacucci+2017b} for this task. This code was specifically designed to simulate the accretion and irradiation processes on high-redshift black hole seeds.

GEMS simulates the accretion flow on to high-redshift black hole seeds, focusing on spatial scales of the order of the Bondi radius $R_B = 2 G M_\bullet / c_s^2(\infty)$ \citep{Bondi1952} of the central source, where $M_\bullet$ is the black hole mass and $c_s (\infty)$ is the sound speed of the unperturbed gas. While the code does not simulate the formation of the seed itself, once the initial mass $M_\bullet (t=t_0)$ is set GEMS is able to compute the output spectrum until complete gas depletion \citep{Pacucci+2017a}. 
In particular, GEMS computes the frequency-integrated radiative transfer through the gas content of the host. The frequency-dependent radiative transfer is performed in a post-processing step with the publicly available spectral synthesis code CLOUDY \citep{Ferland+1998,Ferland+2013}. The final result is the calculation of the time-evolving spectrum emerging from the galaxy hosting the black hole seed.  
The time-dependent data used by CLOUDY includes the following: (i) spatial profiles for hydrogen number density and gas temperature, (ii) the shape of the emission spectrum of the central source of radiation, and (iii) the bolometric luminosity $L_\mathrm{bol}$ of the source. The latter is computed directly from the accretion rate. The SED of the source, ranging from far-infrared to hard X-ray, includes the three, classic, components: (i) a multicolour blackbody, (ii) a power-law, and (iii) a reflected component \citep[see e.g.][for details]{Yue+2013b}.

\begin{figure}
    \centering
    \includegraphics[width=0.45\textwidth]{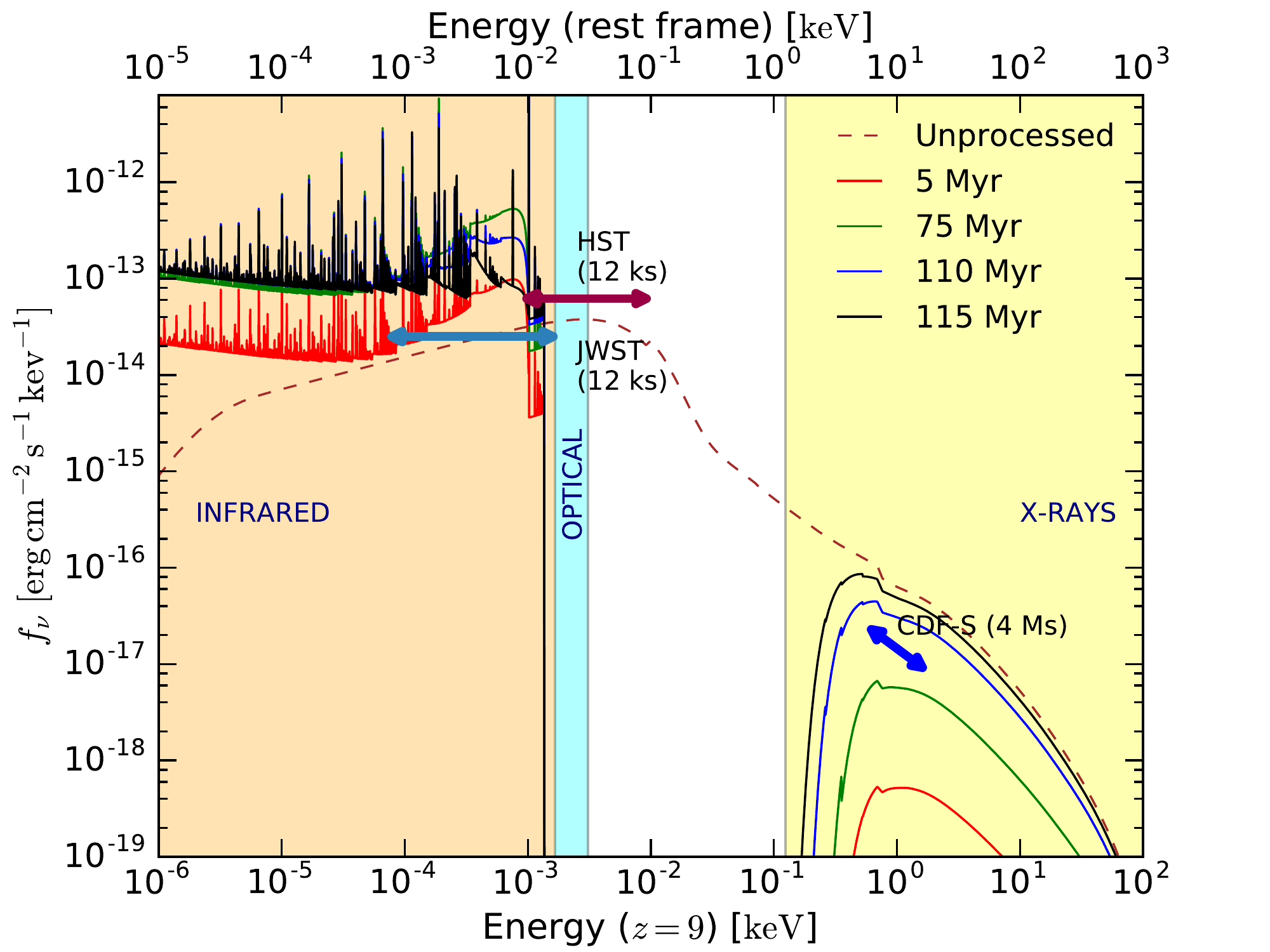}
    \caption{Example of spectral predictions for a typical high-mass black hole seed produced by GEMS. The lines correspond to different times in the evolution of the system (see the legend). The infrared, optical and X-ray bands are shown with shaded regions, while the unprocessed spectrum is reported, at peak luminosity, with a dashed line. The flux limits for future JWST, HST, and CDF-S show that these sources emit a substantial amount of flux in both the infrared and the X-ray wave-lengths.
    \label{fig:dcbh_spectrum}}
\end{figure}

In order to feed initial conditions into the semi-analytic models, the spectral templates for black hole seeds were computed for seeds of 5 different masses ranging from $10^2 \ \mathrm{M}_\odot$ $10^6 \ \mathrm{M}_\odot$, spanning the Pop III remnant to DCBH formation channels \citep{Pacucci+2017a,Woods+2018}.  The templates are computed at different redshifts, to input into the semi-analytic models with the correct spectral properties depending on the cosmic time. With a redshift step $\Delta z=2$, the range of variation spans from $z=6$ \citep[end of the epoch of reionization][]{Kashikawa+2006} to $z=20$ \citep[formation of the first stars][]{Miralda-Escude2003}.  We also explore the effect of the hydrogen column density $N_H$ on the spectral energy distribution \citep{Pacucci+2017b}, and evaluate the effect of values of $N_H$  between $10^{23} \ \mathrm{cm}^{-2}$ (i.e., Compton thin) and $10^{26} \ \mathrm{cm}^{-2}$ (i.e., extremely Compton thick) for these early accreting sources.  We then select the simulation snapshot which maximizes the total  output flux.  An example of a typical spectral template for a $10^5 \ \mathrm{M}_\odot$ black hole seed is shown in Fig. \ref{fig:dcbh_spectrum}. The template extends from the far infrared to the hard X-ray bands.

\begin{figure}
    \centering
    \includegraphics[width=0.45\textwidth]{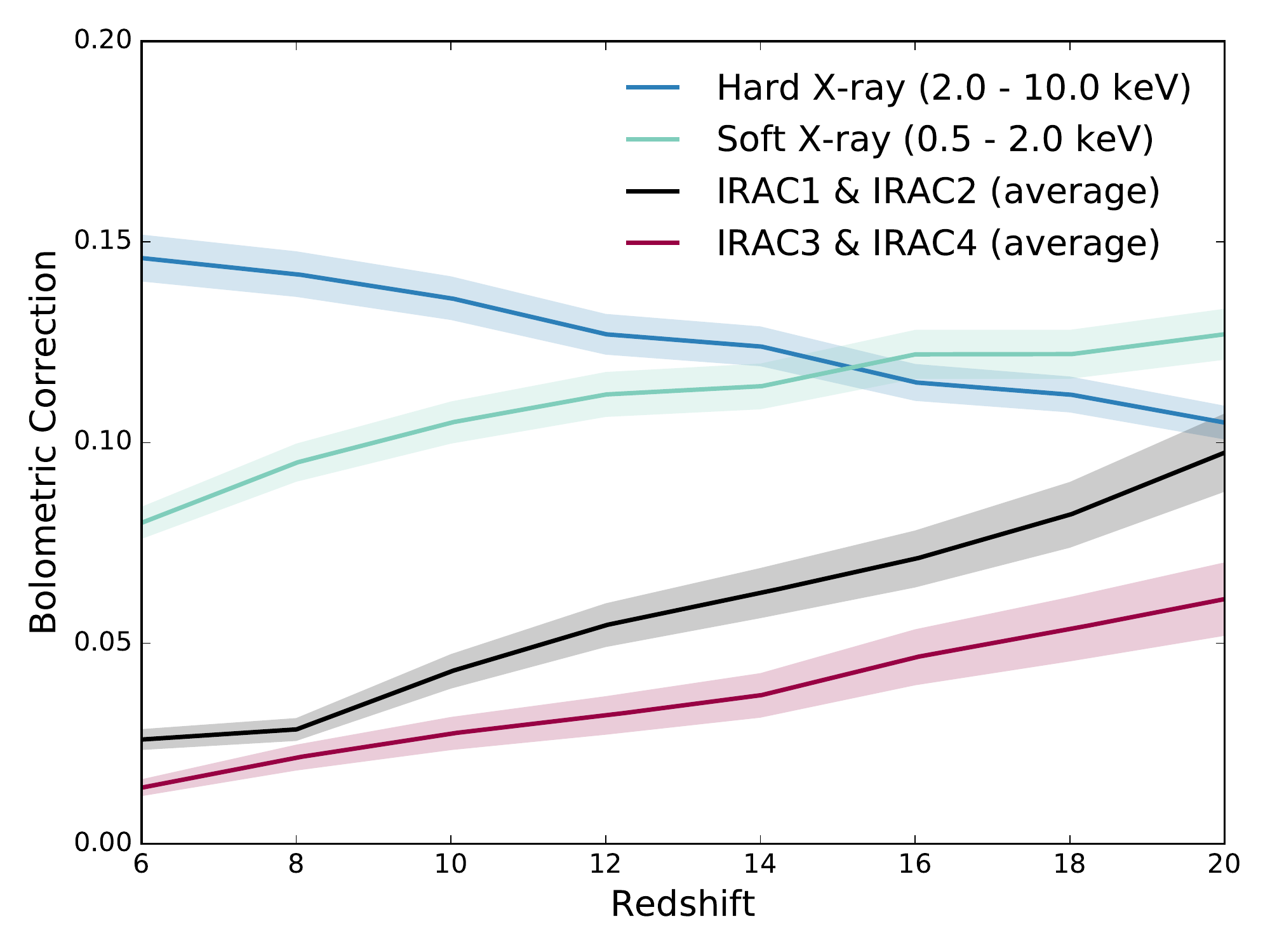}
    \caption{Example bolometric corrections for black hole seeds of $10^{5} \ \mathrm{M}_\odot$ as a function of redshift.  Data for X-ray and NIR bands are shown.
    \label{fig:bolometric_corrections}}
\end{figure}

From these spectral templates, we derive the bolometric corrections as a function of the redshift and mass of the accreting black hole.  These bolometric corrections are needed in order to compute the variation $dF/dz$ of the flux $F$ in the infrared and X-ray bands as a function of the redshift.  The bolometric correction is defined as the ratio between an object's luminosity emitted in some band to its bolometric (total) luminosity.  Fig. \ref{fig:bolometric_corrections} shows the bolometric corrections for a fixed ($10^5 \ \mathrm{M}_\odot$) black hole mass but at different energies with the 1-$\sigma$ uncertainty bands. Overall, the bolometric corrections for infrared bands increase with redshift, with a mean value of $\sim$2 per cent at $z=6$ and reaching $\sim 7$ per cent at $z=20$. The relevance of infrared bands for these sources increases with increasing redshift, confirming the importance of the upcoming James Webb Space Telescope (JWST) mission to search for high-redshift black hole seeds \citep{Natarajan+2017}. The bolometric corrections for X-ray bands are higher, with an average value of $\sim 12$ per cent for both bands. The soft (0.5-2.0 keV) and hard (2.0-10.0 keV) X-ray bands have opposing trends with redshift, due to the fact that the bell-shaped X-ray spectrum shifts with redshift. At $z\sim 15$, the bolometric corrections for soft and hard X-ray bands equal each other at $\sim 12$ per cent of the total luminosity of the source.

\subsubsection{Masking Sources}

\begin{figure}
    \centering
    \includegraphics[width=0.45\textwidth]{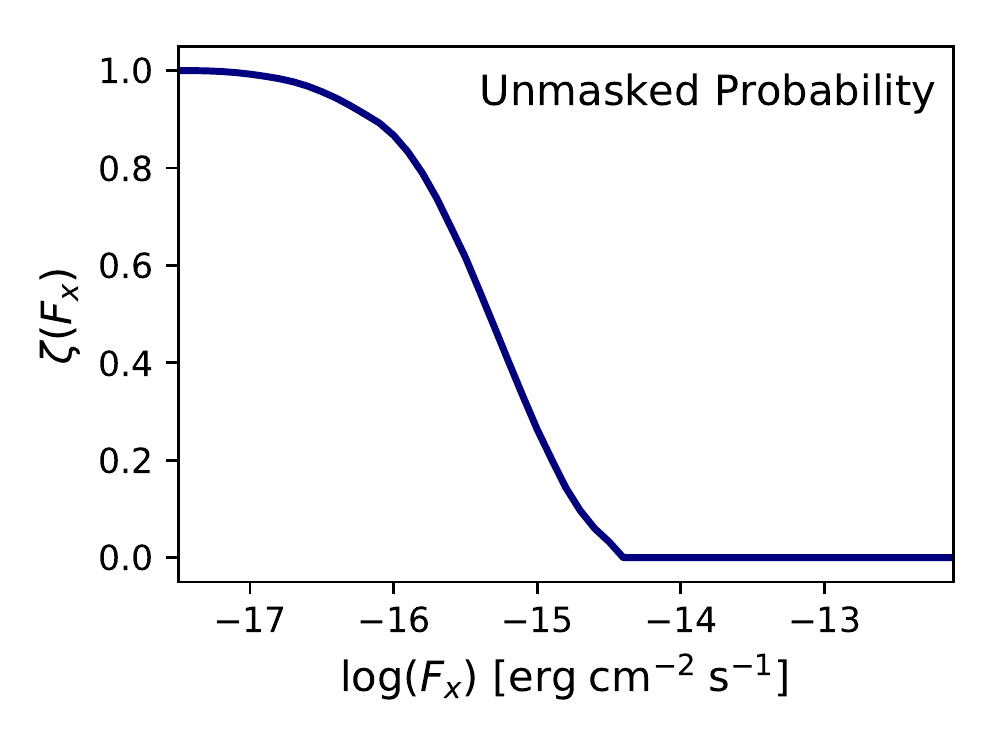}
    \caption{The probability that a source is part of the background, based only on its X-ray flux.  We combine this with IR information to determine the probability of a source being masked.
    \label{fig:unmasked_probability}}
\end{figure}

In the observations, resolved sources are masked in order to sift out the diffuse component of the cosmic backgrounds.  We include the effect that this selection has on the power spectra in our model, removing any accreting MBH with an IR flux brighter than an AB magnitude $\sim 25$ or an X-ray flux brighter than $\sim 10^{-17} \ \mathrm{erg} \; \mathrm{s}^{-1} \; \mathrm{cm}^{-2}$.  The infrared component of the masking is a simple cut on the apparent magnitude at 4.5 $\mu m$, while in fact the X-ray component has a slightly more complicated form due to Chandra's variable sensitivity as a function of off-axis angle and epoch.  Encapsulating these effects, we define the unmasked probability, $\eta (F_x,F_{IR})$, to be

\begin{equation}
    \eta(F_x,F_{IR}) = \begin{cases} 0 & \mathrm{if} \ F_{IR} <25, \\ 
    \zeta(F_x) & \mathrm{otherwise} \\ \end{cases}
\end{equation}

\noindent where $\zeta (F_x)$ is the probability that a source would be undetected in Chandra's AEGIS field.  $\zeta (F_x)$ is reproduced in Fig. \ref{fig:unmasked_probability}.  We have experimented with an IR magnitude cut of 24 instead of 25, and found that this increases the total flux of the derived background by a factor of $\sim 2$.  As we shall show, our main results are insensitive to uncertainties of this level.

\subsection{The Power Spectra of Undetected Sources}\label{sec:clustering}

We begin with the methodology outlined in \citep{Helgason+2014} in order to compute the clustering signal of high-redshift AGN, but make new and critical modifications in order to derive the power spectrum of undetected sources.  Traditional halo occupation distribution (HOD) modelling is based on the number of central and satellite sources as a function of halo mass \citep{Cooray&Sheth2002}.  This implicitly assumes that all accreting MBHs in the model have the same flux (at least at fixed redshift), which is generally untrue.  A generic prediction of almost all MBH assembly models is that more massive haloes host more massive MBHs, and therefore more luminous MBHs cluster more strongly.

Instead of computing the {\it number} of central and satellite sources as a function of halo mass, we compute the {\it flux} attributed to central and satellite galaxies directly from the SAM.  Reformulating in this way avoids issues defining an AGN based on its Eddington ratio or luminosity and allows us to correctly treat AGN at different redshifts.  For example, if there is a fixed flux below which sources are part of the background, the power spectrum in that event probes more luminous sources as redshift increases.

The observed power spectrum can be expressed as a sum of the clustering and shot noise of the emitting population,

\begin{equation}
    P_\mathrm{tot} = P(q) + P_\mathrm{SN},
    \label{eqn:ps_physical_plus_sn}
\end{equation}

\noindent where $q$ is the angular wavenumber in rad$^{-1}$.  Note that the shot noise component is a scale-independent floor.  The first term can be computed by integrating contributions as a function of redshift,

\begin{equation}
    P(q) = \int \frac{H(z)}{c d_c^2(z)} \frac{dF_i}{dz} \frac{dF_j}{dz} P_{i,j} (q d_c^{-1},z) dz,
    \label{eqn:power_spectrum}
\end{equation}

\noindent where $d_c (z)$ is the comoving distance to redshift $z$, $H(z) = H_0 \sqrt{\Omega_M(1+z)^3 + \Omega_\Lambda}$, and $dF_{i,j}/dz$ is the flux gradient of the unmasked sources in observed bands $i,j$.  To proceed, we therefore compute the contribution to the angular power spectrum at fixed redshift intervals, then integrate these pieces.

\subsubsection{Total Flux Per Redshift Interval}

\begin{figure*}
    \centering
    \includegraphics[width=\textwidth]{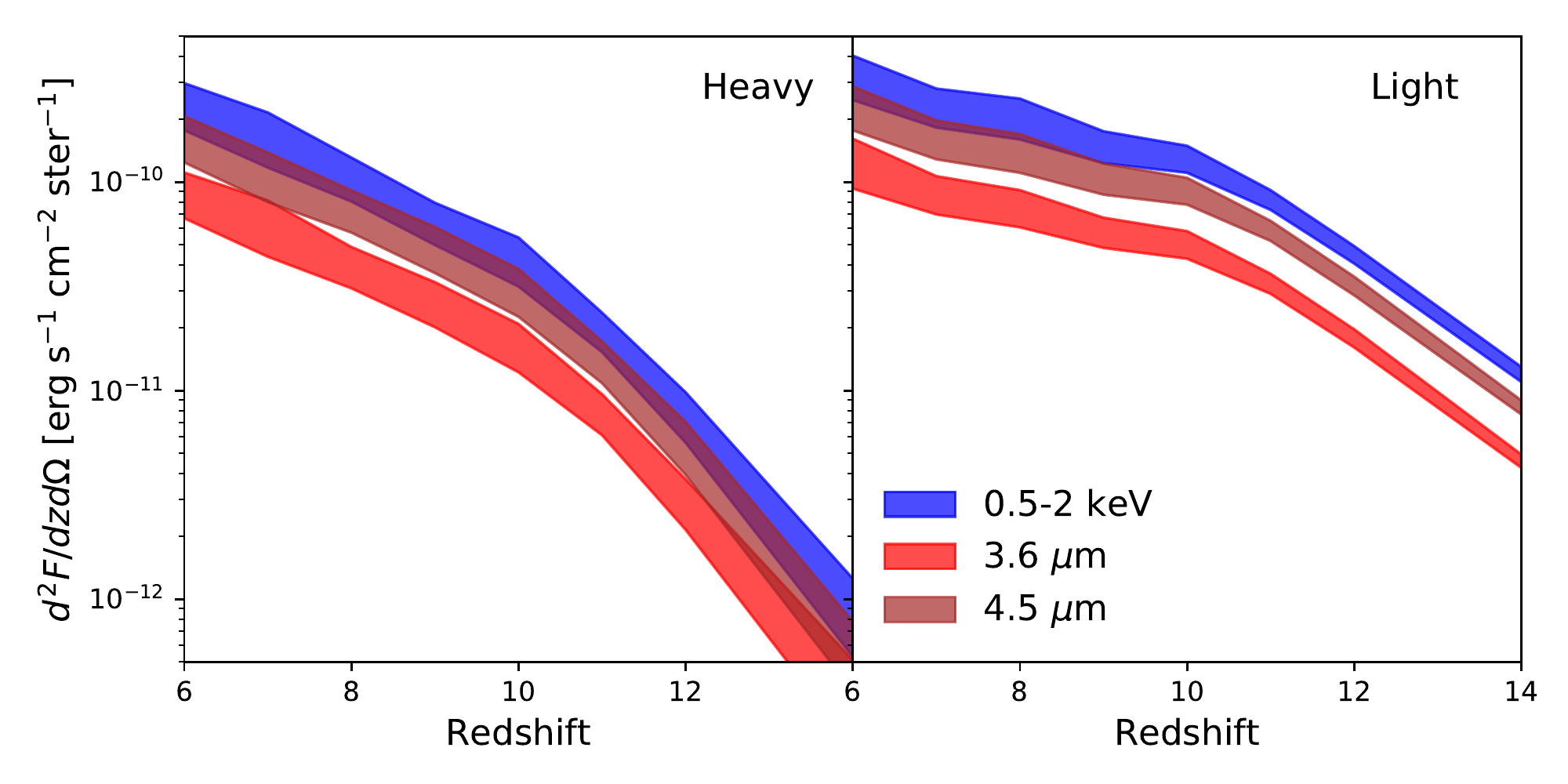}
    \caption{Flux contributions per wavelength interval for light and heavy seeds.  Light seeds produce a factor of 2.0 greater background due to their increased abundance.  Note that the flux contributions in the different bands are all similar.
    \label{fig:flux_gradients}}
\end{figure*}

In this section, we compute the total flux in each relevant band from MBHs as a function of redshift, $dF_{i,j}/dz$ in Equation \ref{eqn:power_spectrum}.  The flux gradient in band $i$ is given by

\begin{equation}
    \frac{dF_i}{dz} = \int \frac{d^3 N}{dF_i d\Omega dz}\eta(F_x,F_{IR}) dF_i .
\end{equation}

\noindent We can express $d^3 N / dF_i d\Omega dz$ as

\begin{equation}
    \frac{d^3 N}{dF_i d\Omega dz} = \frac{d^2 V}{dz d\Omega} \frac{d^2 N}{dF_i dV} .
\end{equation}

\noindent Here, the first factor is the volume-weighting, which comes from standard cosmology,

\begin{equation}
    \frac{d^2 V}{dz d\Omega} = c d_L^2 (1+z)^{-1} \frac{dt}{dz} .
\end{equation}

\noindent The second factor is the flux distribution, calculated directly from the semi-analytic model after applying bolometric corrections.  The results of these calculations are shown in Fig. \ref{fig:flux_gradients} for the soft X-rays, 3.6 $\mu m$, and 4.6 $\mu m$.  Note that these calculations are performed at the observed, not intrinsic, wavelength.  The light seed model produces a factor of 2.0 more background, due to the increased abundance of these seeds compared to the rarer heavy ones.

\subsubsection{Flux-weighted HOD Modelling}

In this section, we compute how the flux at a given redshift is spatially distributed, $P_{i,j} (q d_c^{-1},z)$ in Equation \ref{eqn:power_spectrum}.  In this section, all quantities are computed only for a single redshift slice and for a single band.  We employ the halo occupation distribution (HOD) modelling \citep{Cooray&Sheth2002} technique, but modify from previous approaches by weighting our accreting MBHs by flux.  In this picture, cosmology provides the clustering of dark matter haloes, and the distribution of sources within these haloes modifies the dark matter clustering signal.  

HOD modelling typically begins with the number of centrals and satellites as a function of halo mass as its basis.  Again, this is not appropriate for the clustering of an unresolved background, because this implicitly assumes that AGN luminosities are independent of halo mass.  Instead, we directly compute from our model the total amount of flux associated with centrals and satellites in a given band.

The power spectrum can be written as a sum of the one- and two-halo terms:

\begin{equation}
    P(k) = P^{1h}(k) + P^{2h}(k)
\end{equation}

These represent components arising from the intra- and inter-halo clustering respectively.  In order to solve for these components, we first determine the average AGN flux $F$ in the band that is contributed by the central and satellite populations, weighted by the unmasked probability:

\begin{equation}
    F^c = \langle F^c (M_h,z) \eta(F_x,F_{IR}) \rangle ,
\end{equation}

\noindent and

\begin{equation}
    F^s = \langle F^s(M_h,z) \eta(F_x,F_{IR}) \rangle .
\end{equation}

Since we use Monte Carlo merger trees in order to generate our population, the distinction between central and satellite is simple:  satellites are those sub-haloes which have merged in the merger tree, but have not experienced enough time orbiting their central halo to completely disrupt \citep[the dynamical friction time-scale,][]{Boylan-Kolchin+2008}.  We assume that the light of the satellite population is on average distributed in a Navarro-Frenk-White (NFW) profile \citep{Navarro+1997}.  In practice, the one-halo term is negligible for our study due to the high redshifts we consider, but we include it for completeness in our formalism.  The mean flux density at a given redshift can then be calculated by integrating both components over the halo mass function \citep{Murray+2013}.

\begin{equation}
    \bar{F} = \int \frac{dn}{dM_h} \left[F^c + F^s \right] dM_h .
\end{equation}

The flux-weighted central and satellite biases tell us how the distribution of sources is related to the distribution of dark matter at this epoch.  This can then be computed via,

\begin{equation}
    B^c = \int \frac{dn}{dM_h} \frac{F^c}{\bar{F}} b(M_h,z) dM_h
\end{equation}

\noindent and

\begin{equation}
    B^s = \int \frac{dn}{dM_h} \frac{F^s}{\bar{F}} b(M_h,z) u(k|M_h,z) dM_h
\end{equation}

\noindent where $b(M_h z)$ is the linear halo bias using the ellipsoidal collapse formalism \citep{Sheth+2001}, and $u(k|M_h,z)$ is the Fourier transform of an NFW profile, normalized such that $u(0)=1$ \citep{Scoccimarro+2001}.  This depends on the mass-concentration relation of dark matter haloes, for which we adopt the $z=5$ relation from \citep{Dutton&Maccio2014}.  Finally, having constructed these quantities for some bands $i$ and $j$, we can construct the one-halo component,

\begin{align}
    & P_{i,j}^{1h}(k) = \int \frac{dn}{dM_h} \times \\ 
    &\nonumber \frac{(F^s_iF^c_j + F^s_jF^c_i)u(k|M_h,z) + F^s_iF^s_ju^2(k|M_h,z)}{\bar{F}_i \bar{F}_j} dM_h ,
\end{align}

\noindent which contains terms correlating centrals and satellites as well as satellites with themselves, and the two-halo component,

\begin{equation}
    P^{2h}_{i,j}(k) = P^\mathrm{lin}(k,z)(B^c_i + B^s_i)(B^c_j + B^s_j) ,
\end{equation}

\noindent where $P^\mathrm{lin} (k,z)$ is the linear $\Lambda$CDM power spectrum \citep{Murray+2013}.

\begin{figure}
    \centering
    \includegraphics[width=0.45\textwidth]{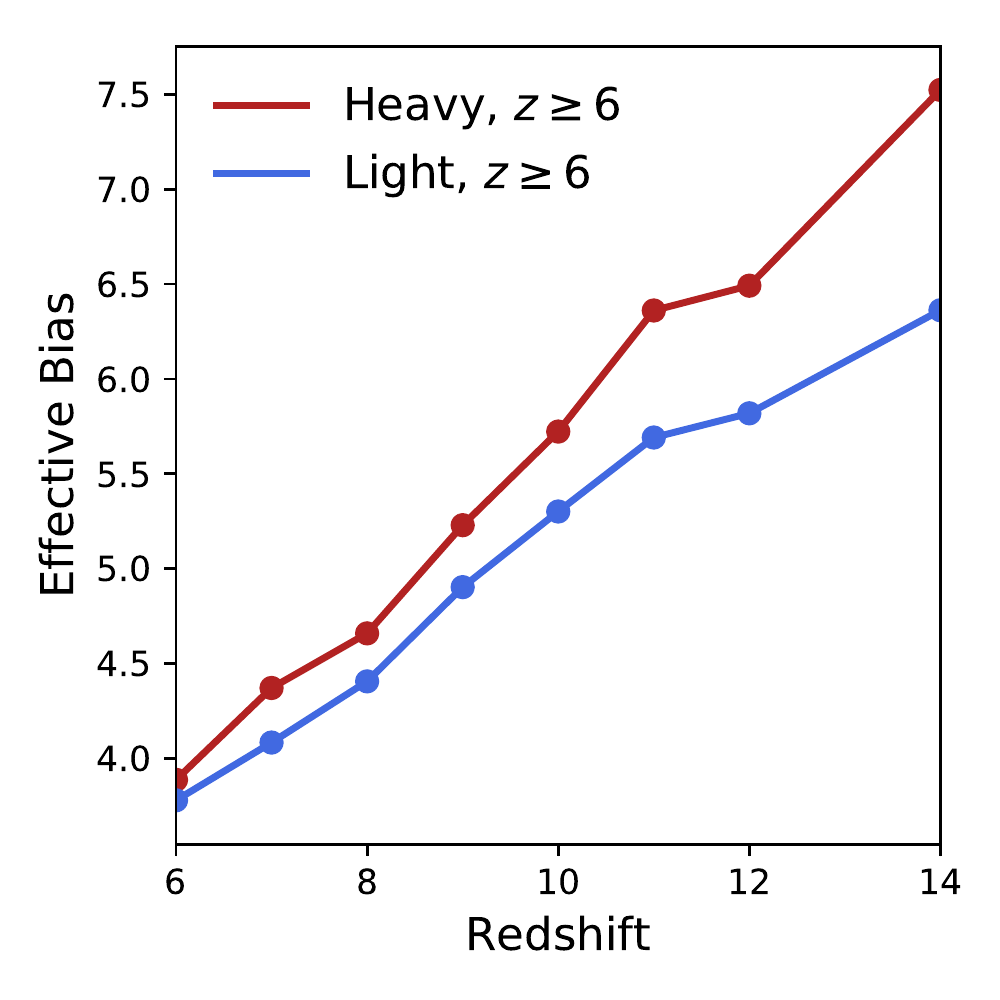}
    \caption{Effective bias for heavy and light seeds.  Heavy seeds produce sources that are slightly more biased, since they tend to occupy higher mass haloes.  Note that since we are dealing with undetected sources, the effective bias is smaller than that of detectable high-redshift quasars or X-ray AGN \citep[e.g.,][]{Cappelluti+2012}.
    \label{fig:effective_bias}}
\end{figure}

While this quantity is not used, it is instructive to compute the ``effective'' bias via

\begin{equation}
    b_\mathrm{eff}(z) = \int \frac{dn}{dM_h} \frac{F^c + F^s}{\bar{F}} b(M_h,z) dM_h .
\end{equation}

\noindent We plot the effective bias for both heavy and light seeds as a function of redshift in Fig. \ref{fig:effective_bias}.  Note that the effective bias of the background at $z=6$ ($\approx 4$) is less extreme than that of quasars at $z\sim 3$ \citep[$\approx 6$ e.g.,][]{Timlin+2018}.  This is because the typical undetected source resides in a much lower mass dark matter halo than a quasar.  Also note that the bias of heavy seeds is higher than that of light seeds, since heavy seeds preferentially occupy higher mass haloes, but this effect is slight.

\subsubsection{Shot Noise}

Computing the shot noise floor is the final missing piece of the power spectrum, $P_\mathrm{SN}$ in Equation \ref{eqn:ps_physical_plus_sn}.  The shot noise power between any given bands $i$ and $j$ is given by 
\begin{equation}
    P_\mathrm{SN} = \int \int F_i F_j \frac{d^2 N}{dF_i dF_j} \eta (F_x,F_{IR}) dF_i dF_j
\end{equation}

\noindent where $d^2 N/dF_i dF_j$ is the distribution of sources with observed fluxes $F_i$ and $F_j$ per unit solid angle.  In practice, we compute $P_{SN}$ at each redshift slice and then integrate over redshift, weighting each component by $d^2 V/dzd\Omega$.  

\subsubsection{One Final Modification: Beam Smearing}

We have now developed the tools to calculate the intrinsic angular power spectrum of hidden/undetected sources on the sky.  Yet in reality, the finite size of the instrument beam suppresses power on small scales.  The final step to computing the observed power spectra is to convolve the signal with the instrument’s point spread function (PSF).  Since we are working in Fourier space, we can simply multiply by its Fourier transform.  We approximate the PSFs of Spitzer and Chandra as Gaussians with full-width half-maximum values of 1.2 arcseconds and 4 arcseconds respectively (the latter being the average over the field of view).  Deviations from Gaussianity are not important for the large scales that we consider in this work.

We do not convolve our model with the mask.  As shown in \cite{Kashlinsky+2012}, mask convolution is only important on small scales, and we exclude $<30$ arcsecond measurements from our analysis.

\section{Power Spectra of Unmasked $z>6$ Black Holes}\label{sec:results_power_spectra}

\begin{figure}
    \centering
    \includegraphics[width=0.45\textwidth]{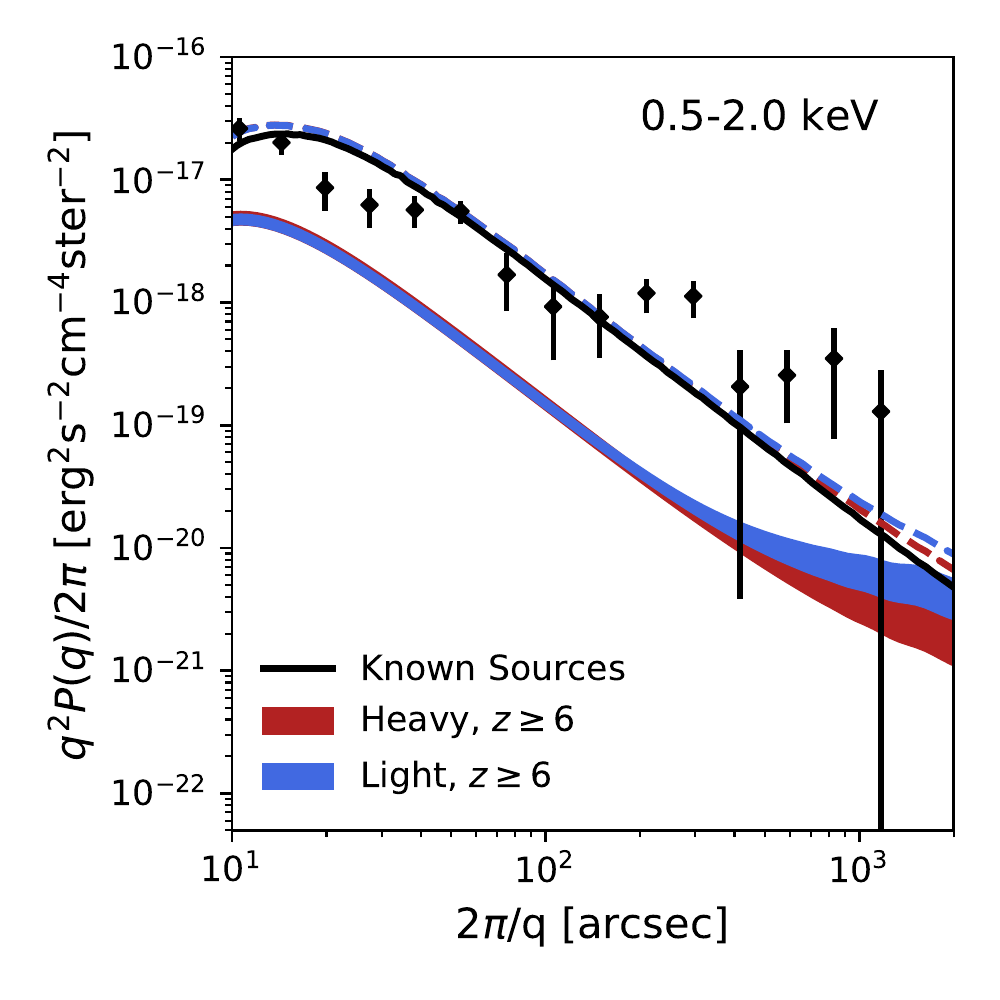}
    \caption{The 0.5-2.0 keV X-ray auto-power spectrum for the background.  Foregrounds for $z < 5$ from \citep{Helgason+2014} are shown as a thin black line, while the colored regions show the contribution from $z>6$ predicted by our model.  Even on large scales, the $z>6$ contribution is not significant.
    \label{fig:xray_auto_sam}}
\end{figure}

\begin{figure*}
    \centering
    \includegraphics[width=\textwidth]{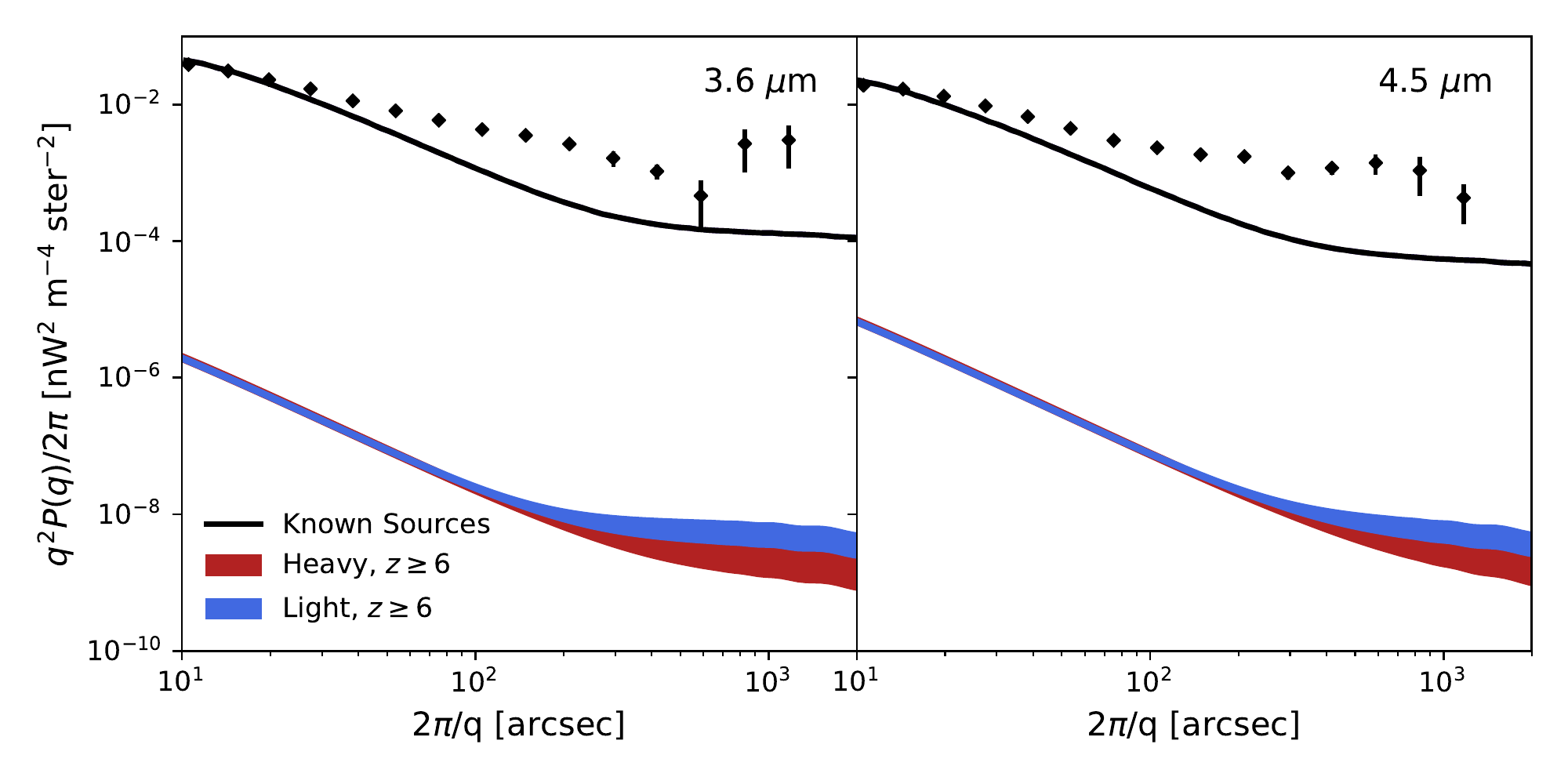} \\
    \includegraphics[width=\textwidth]{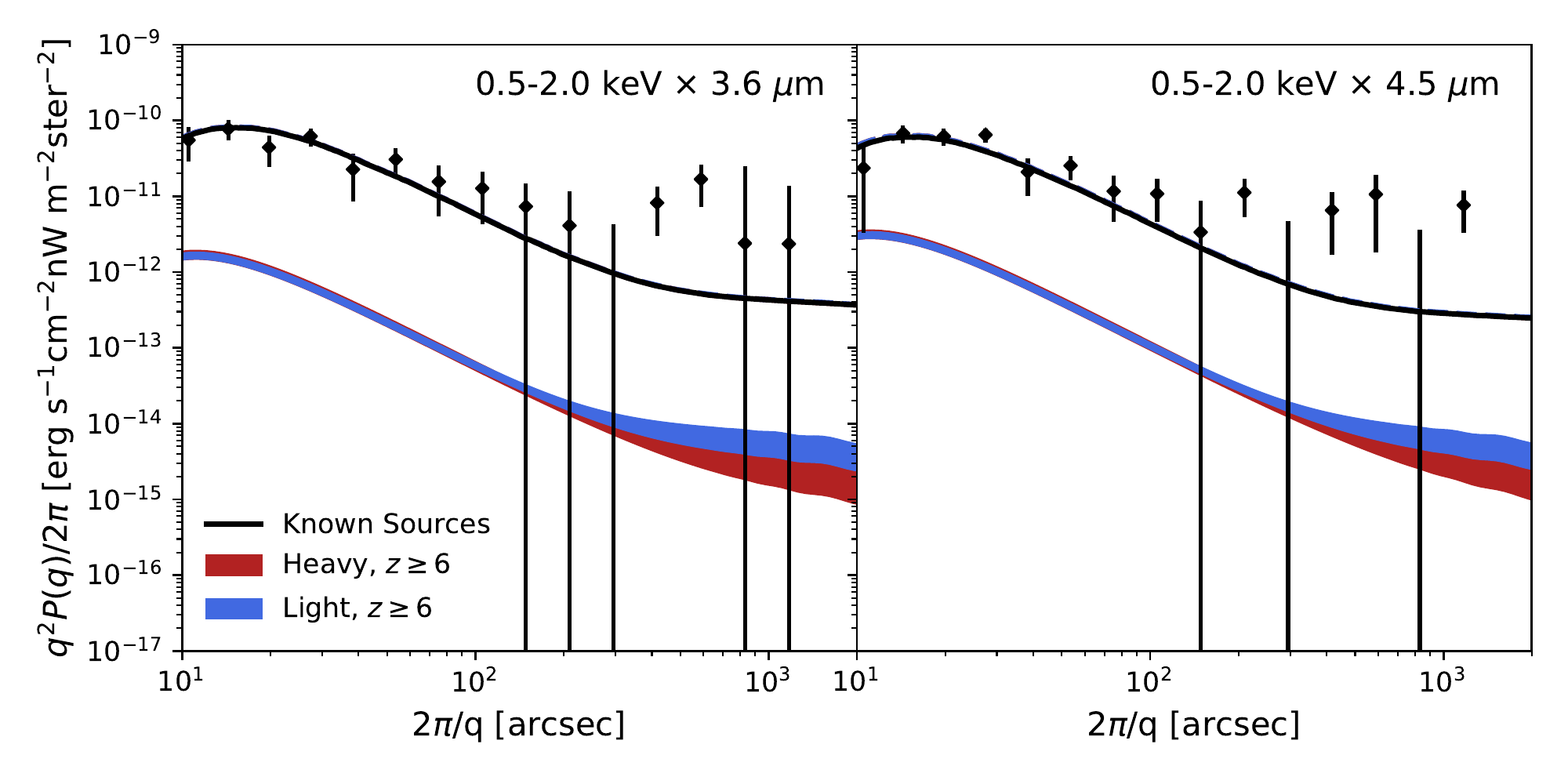}
    \caption{IR power spectra ({\it top}) and IR/X-ray cross power spectra ({\it bottom}) from our model compared to the observations.  The contribution from $z>6$ AGN is shown with the colored regions and is many orders of magnitude below the foregrounds, shown as the thin black line.  The $z>6$ AGN contribution is negligible, even though we have chosen spectra which maximize the infra-red output.
    \label{fig:ir_sam}}
\end{figure*}

We now present the power spectra of unmasked $z>6$ MBHs as predicted by this self-consistent model of MBH assembly, bolometric corrections adopted from hydrodynamical simulations, and a new formalism for computing the power spectrum of undetected sources.  The X-ray auto-power spectrum is shown in Fig. \ref{fig:xray_auto_sam}, while the IR auto-power spectrum and X-ray to IR cross power spectrum are shown in Fig. \ref{fig:ir_sam}.  The power spectrum from the heavy seeding model is shown in red, while that of the light seeding model is shown in blue.  Known foregrounds are shown with a black line \citep{Helgason+2014}, while the data from \citet{Cappelluti+2017a} are shown as black points.

It is immediately clear from the IR auto-power spectrum (Fig. \ref{fig:ir_sam}) that growing $z>6$ MBHs, assembling according to our current understanding of MBH evolution, fall many orders of magnitude short of explaining the excess.  There is very little difference between the power of light and heavy seeds.  Although we found in Fig. \ref{fig:effective_bias} that heavy seeds cluster slightly more strongly, the light seeds actually contribute more power.  This is due to two factors:  (i) light seeds populate more haloes than heavy seeds do, by construction, so there are simply more MBHs that are able to contribute to the background, and (ii) since light seeds are initialized with lower masses than heavy seeds, they must accrete luminously to make up the difference.  The end result is that light seeds produce a factor of 1.7 greater total intensity, and very slightly overtake the heavy seeds.

As seen clearly, this comprehensive model fails dramatically in explaining the excess.  We consider a number of possible explanations, and determine that constraints at lower redshifts on the total mass density of MBHs prevent our model from overgrowing high-redshift MBHs.

\subsection{Higher Bolometric Corrections?} \label{sec:explain_bolometric_corrections}

Is it possible for us to match the data if a larger fraction of the total flux is attributed to the relevant bands?  The answer is no, because each IR band already contains $\approx 2$-$10$ per cent of the total flux.  In the CIB power spectrum, the deficiency of 4 orders of magnitude that we predict corresponds to a deficiency of 2 orders of magnitude in total flux.  However, by adopting a more optimistic bolometric correction, the total flux can only possibly be increased from 2-10 per cent to $\approx 50$ per cent in each IRAC band. 

Previous work has invoked an unusually high obscuring column of $\sim 10^{25} \ \mathrm{cm}^{-2}$ in order to produce the observed ratio between the CIB and CXB fluctuations and prevent premature reionization of the universe \citep{Yue+2013b}.  Our bolometric corrections do not provide as extreme IR-to-X-ray flux ratios because they are taken from the simulation snapshot when the flux is maximized, which is more representative of the time-averaged SED.  Even if simulations are initialized with extremely Compton-thick column densities, the maximum flux is not emitted until the obscuring column has diminished \citep{Pacucci+2015}.  One might then invoke a continuous renewal of the extreme column density, by means of some gas inflow from the outside environment.  In this case, the reduced radiative efficiency would require proportionally more MBH accretion, increasing the tension with the MBH mass density which we explore in subsequent sections.

\subsection{Larger Halo Biases?} \label{sec:explain_halo_biases}

Would it be possible to match the data if a larger bias is assumed?  This is impossible due to the magnitude of the discrepancy. Since the power spectrum is proportional to the square of the bias, the bias would need to be increased by 2 orders of magnitude.  While biases of flux-limited AGN samples tend to be a factor of $\approx 2$ higher than what we predict in Fig. \ref{fig:effective_bias} \citep{Cappelluti+2012}, haloes with biases 2 orders of magnitude higher do not exist in the observable universe.

Note that we do not treat the halo bias as a free parameter, but calculate it directly based on the haloes which host MBHs in our model, which is constrained by the local $M_\bullet-\sigma$ relation and redshift-evolving luminosity functions.  The effective bias reflects the haloes whose MBHs contribute most to the background, excluding ones with MBHs that are luminous enough to be masked.

\subsection{Greater MBH Occupation Fractions?} \label{sec:explain_occupation_fraction}

Is it possible for us to match the data by increasing the total number of MBHs?  We argue that it is not, because our model is constrained by the MBH occupation fraction of low-mass galaxies at $z=0$ \citep{Miller+2015}.  The ability for $z=0$ measurements to constrain $z>6$ predictions underscores the power of our self-consistent modelling.  As shown in \citet{Ricarte&Natarajan2018b}, our heavy seeding prescription is consistent with current constraints based on X-ray emission, while the light seeding prescription optimistically assigns almost every low-mass galaxy a MBH seed. 

The only way to increase the number of MBHs that contribute to the background is to assign additional MBHs to galaxies which already have them.  While we do expect galaxies to host ``wandering'' MBHs \citep{Tremmel+2018}, they could not exist in sufficient quantities to explain this deficit.  The deficit of 2 orders of magnitude in total flux requires that every galaxy by $z=6$ contain 100 times as much mass in wandering MBHs as at the centre of galaxy.  In our model, massive galaxies at $z=6$ already acquire an optimistic black hole-to-total stellar mass ratio of $\approx 1:100$.  This means that {\it every} galaxy would need to accrete as much mass onto MBHs as exists in stars, and most of this would be off-centre.  As we further discuss in \S\ref{sec:explain_black_hole_accretion}, even if this enormous excess mass in wandering MBHs exists, it must somehow be prevented from merging with the central MBH and becoming part of the $z<6$ MBH mass budget.  It is difficult to simultaneously argue that 100 times more mass is accreted onto MBHs, but these masses are also excluded from the centres of galaxies.

\subsection{More Black Hole Accretion?} \label{sec:explain_black_hole_accretion}

Is it possible for us to match the data by simply allowing for greater amounts of MBH accretion?  Here we argue that constraints at lower redshifts do not permit this, and this is the main source of our disagreement with previous studies.  Recall that MBH growth is capped by the $M_\bullet-\sigma$ relation in our model.  It is this cap which limits the rate at which MBHs are allowed to grow on cosmic timescales, not the Eddington ratio distribution, which permits super-Eddington accretion rates in our model.  The $M_\bullet-\sigma$ mass cap regulates the shape of the luminosity function at each epoch (see Appendix \ref{sec:model_validation}), ensures that appropriate masses are obtained as a function of host mass at $z=0$, and determines the black hole mass density in place at $z=6$.

\begin{figure}
    \centering
    \includegraphics[width=0.45\textwidth]{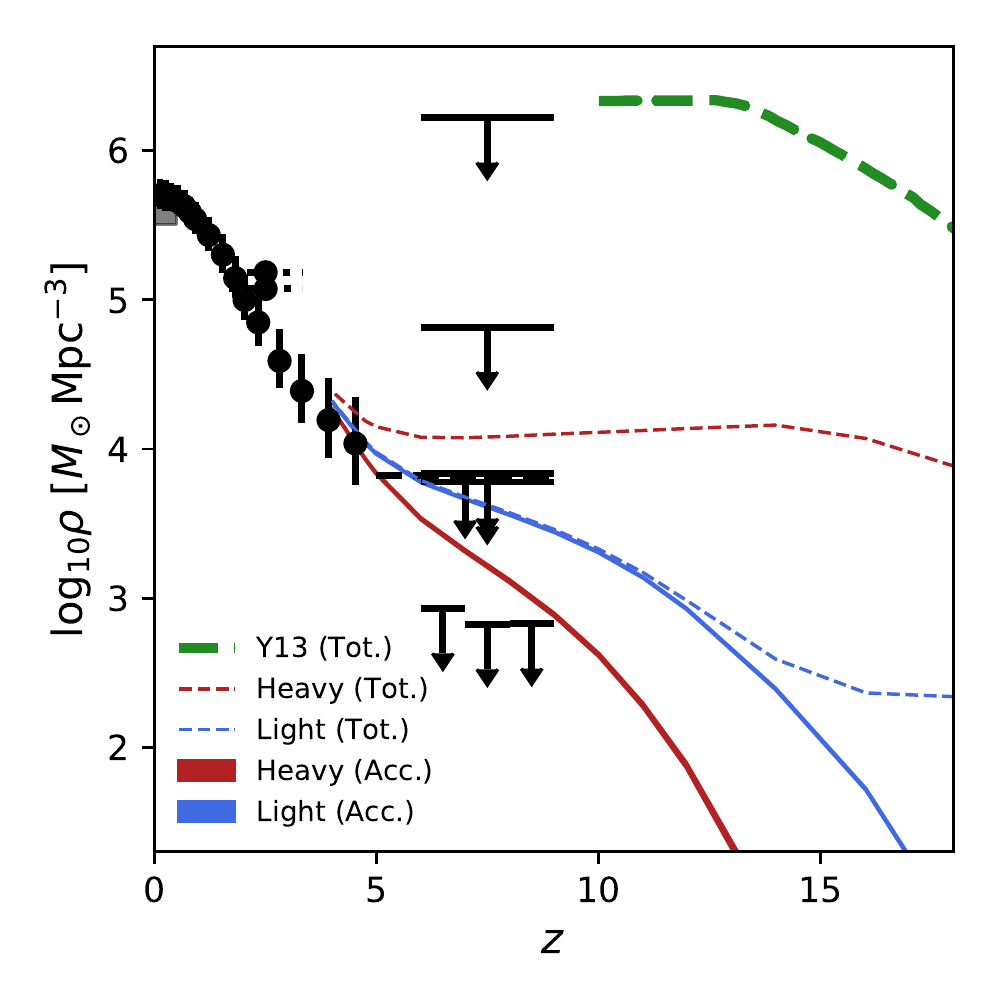}
    \caption{Black hole mass density as a function of redshift.  Filled regions correspond to black hole accretion attributed to luminous accretion, while dashed lines include initial seed masses.  Our model for high-redshift accreting black holes predicts many orders of magnitude less black hole mass density at high redshift than required by the \citet{Yue+2013b} DCBH interpretation, shown in green.  In order of increasing redshift, observational estimates are from \citet{Shankar+2009}, \citet{Hopkins+2007}, \citet{Treister+2009}, \citet{Salvaterra+2012}, and \citet{Cappelluti+2017a}.
    \label{fig:massDensityAndYue}}
\end{figure}

The total amount of mass locked up in MBHs in our model is shown in Fig. \ref{fig:massDensityAndYue}.  We include a large selection of observational estimates based on MBH-host scaling relations at $z=0$, and accretion at higher redshift \citep{Shankar+2009,Hopkins+2007,Treister+2009,Salvaterra+2012,Treister+2013,Cappelluti+2017a}.  The large range of upper limits at $z\sim 6$ is due to different assumptions regarding the fraction of accreted black hole mass that is converted into observable X-ray emission.  The constraints from \citet{Shankar+2009,Hopkins+2007} should be considered more robust, however, as these are based on local empirical relationships between MBHs and their hosts and bolometric luminosity functions respectively.  We plot the total amount of mass locked up in MBHs with dashed lines (this includes seed masses), and the accreted portion of the mass locked up on MBHs as a solid region.  In green, we also plot the total integrated mass density required by the  \citet{Yue+2013b} model which explains the missing fluctuations with early MBH growth.  We also provide these mass densities at $z=6$ in Table \ref{tab:massDensities}.  Only $10^{3-4} \ \mathrm{M}_\odot \; \mathrm{Mpc}^{-3}$ is in place at this epoch in our model, while the \citet{Yue+2013b} interpretation requires $> 10^6 \ \mathrm{M}_\odot \; \mathrm{Mpc}^{-3}$ to be in place by $z=10$.

\begin{table}
\begin{tabular}{lll}
\hline
Seed Type & $\log \rho_\mathrm{tot}(6) \ [\mathrm{M}_\odot \; \mathrm{Mpc}^{-3}]$ & $\log \rho_\mathrm{acc}(6) \ [\mathrm{M}_\odot \; \mathrm{Mpc}^{-3}]$ \\
\hline
Heavy     & $3.9755 \pm 0.0098$                                        & $3.429 \pm 0.007$                                          \\
Light     & $3.685 \pm 0.003$                                          & $3.674 \pm 0.003$                                         \\
\hline
\end{tabular}
\caption{Integrated mass densities of central black holes at $z=6$ predicted by the semi-analytic model.  The total mass density (left) includes seed masses, while the accreted mass density (right) does not. \label{tab:massDensities}}
\end{table}

In the \citet{Yue+2013b} model, so much mass is accreted onto MBHs that the integrated mass density at $z=10$ already exceeds the mass density at $z=0$.  Such large mass densities are required simply to produce enough flux, as explored in detail in \citet{Helgason+2016}.  As discussed in \S\ref{sec:explain_occupation_fraction}, this leaves no room for MBHs to grow at lower redshifts.  These accreting MBHs would somehow need to simultaneously accrete far more efficiently than expected, and yet remain displaced from the centres of galaxies throughout all of cosmic time.  Note that burying these sources in large columns {\it does} permit such a model to agree with the intensity of the X-ray background \citep{Cappelluti+2017a}---however, it is bolometric luminosity density measurements at lower redshift \citep{Hopkins+2007} and local scaling relations \citep{Shankar+2009} which rule out this interpretation.

\section{Discussion}\label{sec:discussion}

The comprehensive model developed here fails by many orders of magnitude to reproduce the missing CIB/CXB fluctuations with $z>6$ accretion.  A successful model would need to produce about 2 orders of magnitude more total flux in the IR.  The main reason that our results disagree with previous attempts to explain this excess with accreting black holes \citep{Yue+2013b} is that this previous work requires more mass locked up in MBHs at $z=10$ than there exists at $z=0$, which is not permitted in our models of self-consistent growth over cosmic time.  In this section, we explore potential alternative models.  In \S\ref{sec:moderatez_agn}, we consider missing AGN at lower redshift.  In \S\ref{sec:foreground}, we consider a missing Galactic foreground component.

\subsection{Can we be missing buried moderate-redshift AGN?} \label{sec:moderatez_agn}

\begin{figure}
    \centering
    \includegraphics[width=0.45\textwidth]{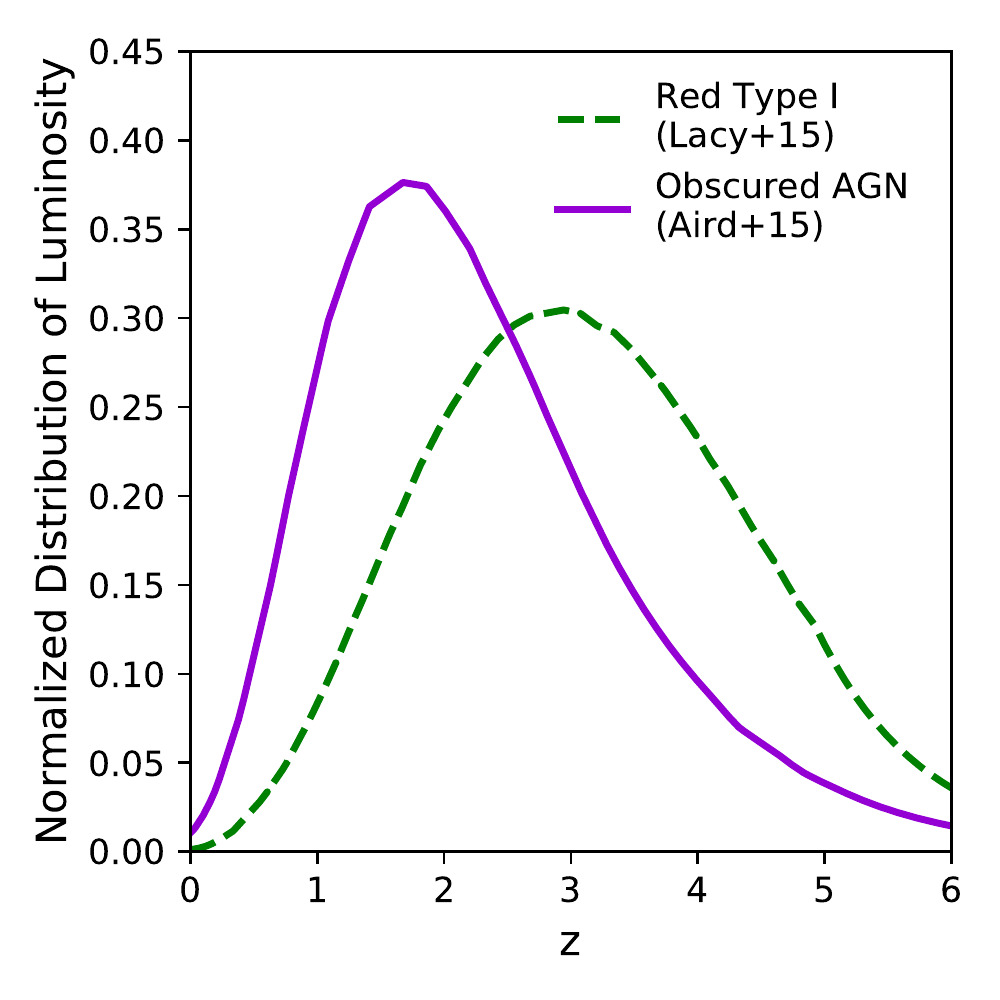}
    \caption{Redshift distributions of hypothetical moderate-redshift AGN.  For plausible distributions, we assume that their fluxes are distributed in redshift according to either red type I AGN from \citet{Lacy+2015} or obscured AGN from \citet{Aird+2015}.
    \label{fig:redshift_distributions}}
\end{figure}

We now explore the existence of missing AGN not at $z>6$, but rather at $z\sim 2-3$, unaccounted for in our census of SMBH growth. Missing SMBH accretion can be more easily accommodated at these epochs than at $z>6$ due to the amount of SMBH mass already in place.  Interestingly, population synthesis models that reproduce the observed spectral shape of the CXB require the existence of a population of Compton-thick actively accreting black holes that may account for 30-50 per cent of SMBH growth \citep{Treister+2004,Ballantyne+2006,Gilli+2007,Aird+2015,Ananna+2019}.  Some evolutionary models of AGN triggering suggest that MBH growth may begin in a highly obscured phase before blowing away the obscuring column and becoming observable as a traditional AGN \citep{Hopkins+2008,Alexander&Hickox2012,Blecha+2018}.  Such an evolutionary phase before each AGN episode could be accommodated in a self-consistent model by modifying the average radiative efficiency of an accretion disk \citep{Soltan1982}.

We hypothesize that the flux from such missing buried sources could be distributed as a function of redshift according to the luminosity densities of either obscured AGN \citep{Aird+2015} or red type I quasars \citep{Lacy+2015}, whose redshift distributions are shown in Fig. \ref{fig:redshift_distributions}.  Since these redshift distributions peak around $z\sim 2$ and $z\sim 3$, we label these hypothetical missing populations AGN\_z2 and AGN\_z3 respectively.  We further assume that they occupy a characteristic halo mass of $10^{13} \ \mathrm{M}_\odot$, the typical halo mass of X-ray selected AGNs \citep{Cappelluti+2012}.  This characteristic halo mass provides an optimistically high bias factor, since such high mass haloes are likely to host galaxies which would be masked.   We do not include a satellite contribution to the flux of this population, and set the one-halo term to zero.

\begin{table*}
\begin{tabular}{lllll}
\hline
        & $\log (F_{\mathrm{tot},3.6})$ & $\log (F_{3.6}/F_{4.5})$ & $\log (F_x/F_{3.6})$ & $\log C_\mathrm{X-IR}$ \\
\hline
AGN\_z2 & $-6.317^{+0.010}_{-0.011}$ & $-0.063^{+0.014}_{-0.013}$ & $-2.09^{+0.04}_{-0.46}$ & $-0.6^{+0.2}_{-0.7}$ \\
AGN\_z3 & $-6.438^{+0.010}_{-0.011}$ & $-0.064^{+0.014}_{-0.013}$ & $-2.08^{+0.05}_{-0.38}$ & $-0.6^{+0.2}_{-0.6}$ \\
\hline
\end{tabular}
\caption{Best fit parameters for an unknown extra-galactic population of AGN with a characteristic host halo mass of $10^{13} \ \mathrm{M}_\odot$.  $F_{\mathrm{tot},3.6}$ has units of $\mathrm{erg} \; \mathrm{s}^{-1} \; \mathrm{cm}^{-2} \; \mathrm{ster}^{-1}$. Note that the data require an extreme 1:100 flux ratio between the soft X-ray and the 3.6 $\mu m$ bands.  \label{tab:background_parameters}}
\end{table*}

\begin{table}
\begin{tabular}{llll}
\hline
          & $\chi^2$ & $\chi^2/\mathrm{DOF}$ & BIC  \\
\hline
Power Law & 63.6     & 1.27                  & 83.6 \\
AGN\_z2   & 1140     & 22.4                  & 1160 \\
AGN\_z3   & 1090     & 21.5                  & 1110 \\
\hline
\end{tabular}
\caption{Fitting metrics for various alternate models.  Generic extra-galactic populations do not provide good fits due to the gentle slope in the CIB data on arcminute scales.  A simple power law is highly preferred over an extra-galactic population on the basis of the Bayesian Information Criterion (BIC). \label{tab:fittingMetrics}}
\end{table}

Nevertheless, we use these optimistic assumptions to simultaneously fit all power spectra for 4 free parameters: the total amount of isotropic flux in the IRAC1 band, flux ratios to the IRAC2 and soft X-ray bands, and the coherence between the CIB and CXB.  Best fit values are provided in Table \ref{tab:background_parameters}, with errors are estimated using the Markov Chain Monte-Carlo package emcee \citep{Foreman-Mackey+2013}.  Values for $\chi^2$, reduced $\chi^2$, and the Bayesian Information Criterion (BIC) are provided in Table \ref{tab:fittingMetrics}.  The BIC is a useful metric for comparing models with different numbers of parameters \citep{Schwarz1978}.  Additional details of this optimization process can be found in Appendix \ref{sec:fitting}.

\begin{figure}
    \centering
    \includegraphics[width=0.45\textwidth]{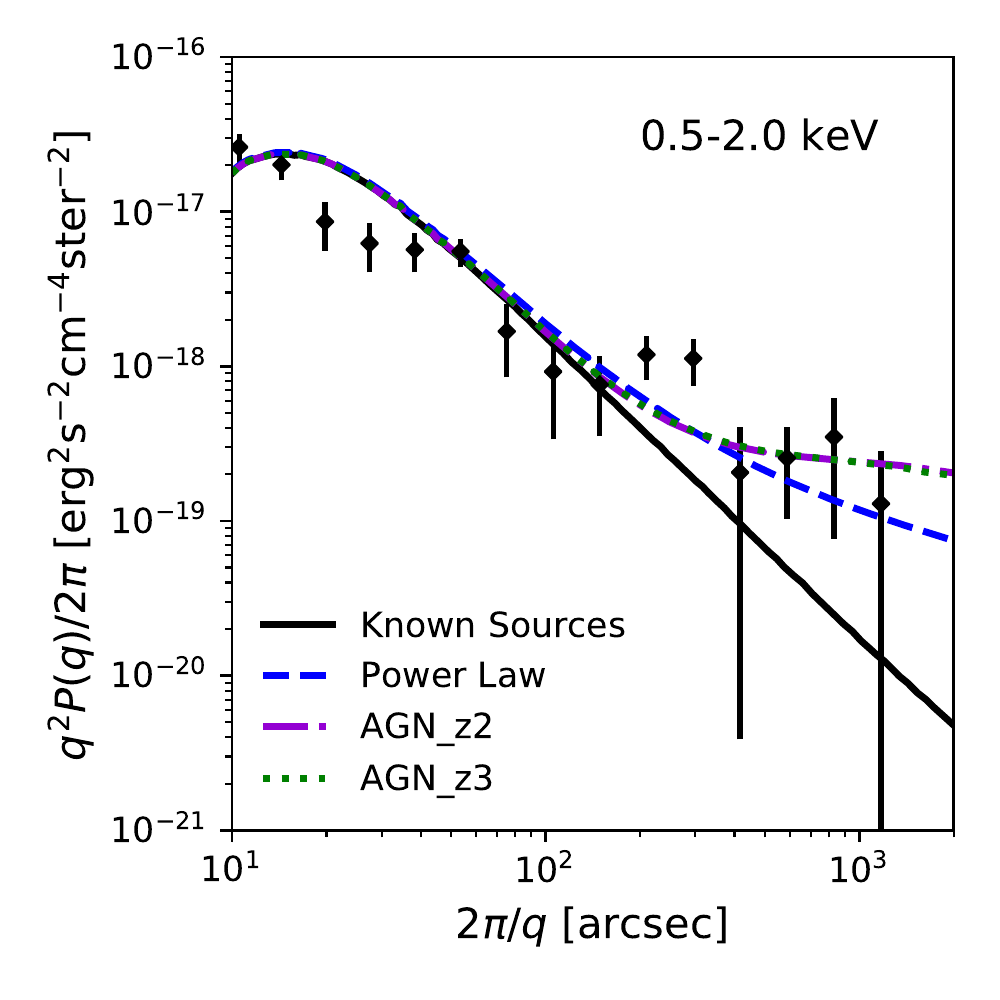}
    \caption{Soft X-ray power spectra for best-fit hypothetical sources.  These correspond to a simple power-law fit, missing AGN populations peaking at $z\sim 2$, and missing AGN populations peaking at $z\sim 3$.
    \label{fig:xray_auto_hypothetical}}
\end{figure}

\begin{figure*}
    \centering
    \includegraphics[width=\textwidth]{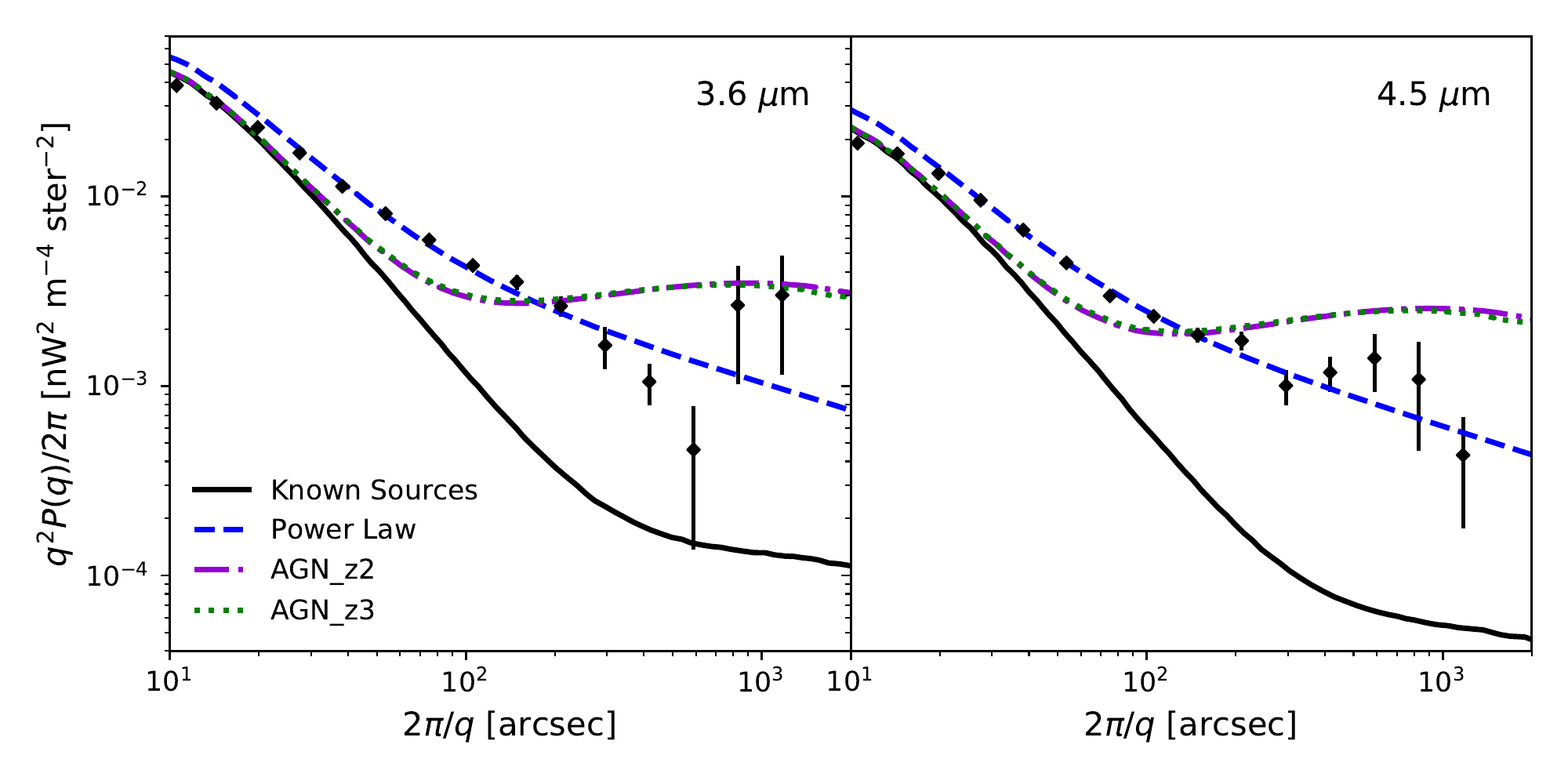} \\
    \includegraphics[width=\textwidth]{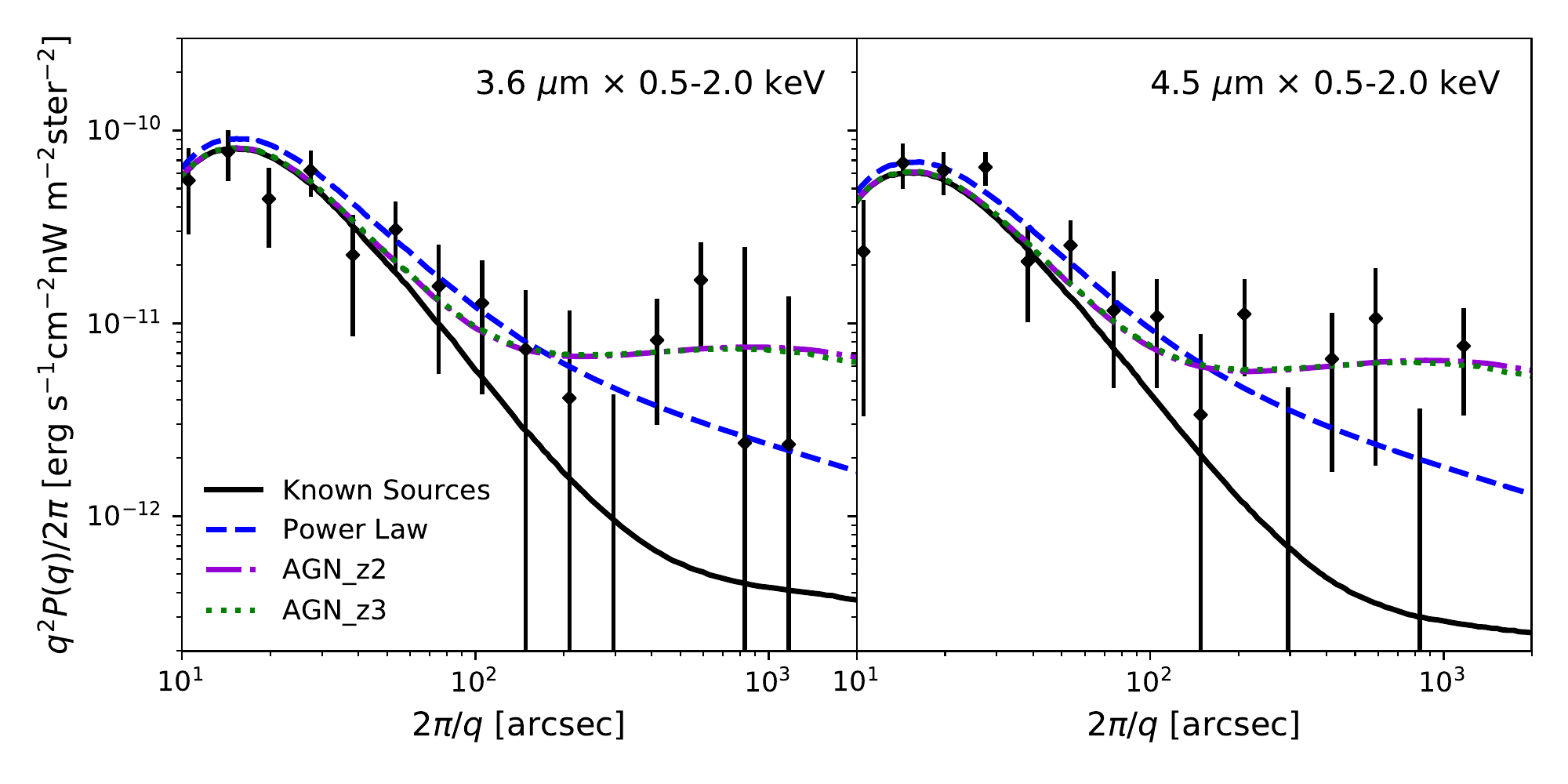}
    \caption{Best fit infrared auto-power spectra ({\it top}) and cross-power spectra ({\it bottom}) for hypothetical sources peaking at $z\sim 2$ and 3.  A generic extra-galactic AGN population struggles to reproduce the shape of the missing fluctuations in auto-power spectra.  A simple power law is highly preferred by the data. 
    \label{fig:ir_hypothetical}}
\end{figure*}

Power spectra for the best fit hypothetical populations are shown in Figs. \ref{fig:xray_auto_hypothetical} and \ref{fig:ir_hypothetical}.  We find that these generic extra-galactic populations also provide a poor fit to the data, with $\chi^2/DOF \sim 20$.  The incompatibility of this model with the data is evident in Fig. \ref{fig:ir_hypothetical}, in the computed auto-power spectrum of the CIB.  The two-halo term of dark matter clustering provides too much power on degree scales relative to arcminute scales, and the best fit attempts to compensate by undershooting the data on small scales and overshooting on large scales.

In addition, as with \citet{Yue+2013b}, the best-fit isotropic flux value translates to an amount of mass in tension with the local black hole mass density.  Following the “Soltan argument” \citep{Soltan1982}, allowing 10 per cent of the bolometric flux to be radiated in each of the IR bands, and assuming a typical radiative efficiency of 0.1, we find that this extra flux would correspond to an extra black hole mass density of $10^{6.0} \ \mathrm{M}_\odot \; \mathrm{Mpc}^{-3}$ for either model, once again in excess of current estimates of the $z=0$ total mass density \citep{Shankar+2009}.  (For more details, see Appendix \ref{sec:soltan}.)  While more informed priors in the fitting process could reduce this value, this would most likely come at the expense of even worse fits to the data.

In principle, the underproduction of power at sub-arcminute scales could be alleviated with the addition of a one-halo term to the power spectrum.  We therefore experimented with a model in which {\it all} of the flux is attributed to satellites distributed according to the appropriate NFW profile for a given mass and redshift.  We found that this modification could boost the power on 10 arcsecond scales by up to 50 per cent for standard halo concentrations.  This increases to up to a factor of 4 if a much higher concentration parameter of 50 is assumed, which some authors suggest is a value of the NFW concentration parameter required to match small-scale AGN clustering \citep{Hennawi+2006,Lin&Mohr2007,Kayo&Oguri2012,Eftekharzadeh+2019}.  However, even in the latter case, the one-halo term is buried beneath the foreground of known sources at these scales.

In addition, the spectral properties of the fluctuations also do not favor an obscured AGN scenario at moderate redshift.  Even obscured AGN do not typically emit strongly at 1 micron, which in fact is a good band to probe the stellar continuum \citep[e.g.,][]{Collinson+2017,Hickox&Alexander2018}.  Reprocessed emission usually emerges in the far infrared, where the optical depth falls.  In addition, a recent analysis of the spectrum of the fluctuations favors unobscured AGN over an obscured population \citep{Li+2018}.  This surprising result is not predicted by any current explanations of the CIB excess.

\subsection{Are we missing a Galactic foreground?} \label{sec:foreground}

The power spectrum of the fluctuations resembles a simple power law much more than the shape of the extra-galactic two-halo term.  In addition, the spectra resemble a Rayleigh-Jeans tail in the NIR and unobscured black hole accretion in the X-ray.  Guided by these facts, we explore the hypothesis of a new Galactic foreground of warm gas and dust spatially distributed separately from colder Galactic cirrus.  We show that although the NIR data would prefer a simple power law that may be due to a Galactic component, reflected Galactic X-ray light is insufficient to explain the coherence with the CXB.

Observationally, it has been difficult to estimate the contribution from Galactic cirrus to the NIR fluctuations, as discussed in detail in \citet{Kashlinsky+2018}.  The main method with which to assess its contribution has been to attempt to find cross-correlations with bands where cirrus is known to be dominant.  Using Spitzer data, \citet{Kashlinsky+2012} find marginal cross-correlations with the NIR and 8 micron emission, while \citet{Matsumoto+2011} find no cross-correlation with longer wavelength 100 micron maps with AKARI.  In contrast, \citet{Thacker+2015} do report a cross-correlation at longer 250, 350, and 500 micron wavelengths, but \citet{Kashlinsky+2018} argues that the claimed signal may be derived from known Galactic populations.  At the shorter wavelengths probed by HST, \citet{Mitchell-Wynne+2015} observe a clear power-law on scales $> 1'$, which they attribute to diffuse Galactic light.  \citet{Zemcov+2014} report a cross-correlation between 1.1 and 1.6 micron fluctuations with the 3.6 micron fluctuations, which they attribute to IHL.  However, \citet{Yue+2016} argue that this cross-correlation is due to diffuse Galactic light based on its power-law behaviour reported by \citet{Mitchell-Wynne+2015}. 

\begin{table*}
\begin{tabular}{lllll}
\hline
$\log P_{1000}$            & $\alpha$                & $\log (F_{4.5} / F_{3.6})$ & $\log (F_x/F_{3.6})$          & $\log C_\mathrm{X-IR}$ \\
\hline
$-20.30^{+0.038}_{-0.058}$ & $-1.47^{+0.04}_{-0.03}$ & $-0.105^{+0.010}_{-0.010}$ & $-1.9778^{+0.0002}_{-5.6118}$ & $-0.7^{+0.3}_{-3.8}$  \\
\hline
\end{tabular}
\caption{Best fit parameters for an unknown foreground component.  Note that while a flux ratio between the CXB and CIB of $\sim 0.01$ is preferred, the large error bars on large scales in this dataset allow for little cross-power between the CIB and CXB.  $P_{1000}$ has units of $\mathrm{erg}^2 \; \mathrm{s}^{-2} \; \mathrm{cm}^{-4} \; \mathrm{ster}^{-1}$. \label{tab:powerLaw_parameters}}
\end{table*}

To explain the coherence with the CXB, we hypothesize that this foreground reflects Galactic X-ray emission, as speculated by \citet{Cappelluti+2017b}.  We show, however, that such reflected emission is too faint for to plausibly explain the cross-correlation with the CXB.  We first perform a simple power-law fit to the missing source of the fluctuations, $P(q) \propto q^a$.  The free parameters of this model are P1000, which is the value of the power spectrum at $q=1000 \ \mathrm{rad}^{-1}$ (or $0.36^\circ$), the slope $\alpha$, the flux ratios between different bands, and the coherence between the CIB and CXB.  Best fit values are shown in Table \ref{tab:powerLaw_parameters}.

The best fit model for a simple power law fit to the CIB and CXB has a power of $5.6 \times 10^{-25} \ \mathrm{erg}^2 \; \mathrm{s}^{-2} \; \mathrm{cm}^{-4} \; \mathrm{ster}^{-1}$ in the soft X-rays at $q=1000 \ \mathrm{rad}^{-1}$.  The root-mean-square flux of fluctuations at this scale is then equal to $(q^2P(q)/2\pi)^{1/2}$ \citep{Kashlinsky&Odenwald2000}, which yields a value of $3.0 \times 10^{-10} \ \mathrm{erg} \; \mathrm{s} \; \mathrm{cm}^{-2} \; \mathrm{ster}^{-1}$.  We equate this to the minimum amount of X-ray flux required by our hypothetical Galactic foreground.  It is possible that only 20 per cent of this value is required, which is the measured coherence between the CIB and CXB.

We approximate the geometry of the Milky Way’s X-ray emitting population to be an infinite disk, above which lies a scattering screen of column depth $N_H \sim 10^{20} \ \mathrm{cm}^{-2}$, as is typical for the observational data derived from the four fields considered in this work.  \citet{Grimm+2002} determine that the total 2-10 keV luminosity of X-ray binaries in the Milky Way is approximately $2-3 \times 10^{39} \  \mathrm{erg} \; \mathrm{s}^{-1}$.  For an order-of-magnitude estimate, we assume that the 0.5-2.0 keV flux is the same and distribute these sources uniformly in a disk of radius 15 kpc, allowing us to estimate the soft X-ray flux of the Milky Way as follows:

\begin{equation}
    F_\mathrm{MW} \sim L_\mathrm{MW} \frac{1}{\pi (15 \ \mathrm{kpc})^2}\frac{1}{4\pi} \sim 3\times 10^{-8} \ \mathrm{erg} \; \mathrm{s}^{-1} \; \mathrm{cm}^{-2} \; \mathrm{ster}^{-1} .
\end{equation}

The fraction of this flux which encounters the scattering screen and is reflected back can be calculated on the basis of its optical depth, $\tau = N_H \sigma_T$, where $\sigma_T$ is the Thomson cross-section.  This yields

\begin{equation}
    F_\mathrm{ref} = F_\mathrm{MW} (1 - e^{-N_H \sigma_T}) ,
\end{equation}

\noindent which for our estimated value of $F_\mathrm{MW}$ and $N_H= 10^{20} \ \mathrm{cm}^{-2}$ results in 

\begin{equation}
    F_\mathrm{ref} \sim 2 \times 10^{-12} \ \mathrm{erg} \; \mathrm{s}^{-1} \; \mathrm{cm}^{-2} \; \mathrm{ster}^{-1} .
\end{equation}

This falls two orders of magnitude short of producing the minimum amount of flux required, which would need a column of at least $\sim 10^{22} \ \mathrm{cm^{-2}}$.  Consequently, reflected Galactic X-ray emission is not a plausible mechanism for generating these coherent CXB fluctuations.

\section{Conclusion}\label{sec:conclusion}

The CIB exhibits unexplained fluctuations on arcminute scales and higher that cross-correlate with CXB fluctuations.  This has motivated some authors to explain the origin of these fluctuations as arising from early, obscured MBH growth \citep{Yue+2013b,Yue+2016}.  For the first time, we examine this hypothesis using a self-consistent semi-analytic model of black hole populations accreting over cosmic time that are capable of incorporating observed constraints at lower redshifts, namely local MBH-to-host relations and bolometric luminosity functions.  We combine these calculations with bolometric corrections from hydrodynamical simulations.  By deriving a new formalism for the power spectrum of undetected sources, we then compute the power spectrum expected due to $z>6$ MBH growth.

We find that $z>6$ MBH accretion cannot explain the missing fluctuations due to the enormous amount of growth that would be acquired at such early epochs.  Previous attempts to explain these fluctuations using early MBH accretion require more mass locked up in MBHs at $z=10$ than there exists at $z=0$.

We then examine alternate solutions to explain the observed excess.  First, we hypothesize that buried AGN at more moderate redshifts, missing from our current census, are responsible.  We find that even an optimized model for an AGN origin at low-redshift produces a poor fit on arcminute scales while simultaneously overproducing the local MBH mass density.  Then, we explore the hypothesis that a new foreground of Galactic dust might produce the missing IR fluctuations by emission and the coherent X-ray fluctuations by reflection.  We find that the Galaxy produces too little X-ray flux for this to plausibly explain the coherence between the two backgrounds.

We decisively rule out accreting MBHs as the sources of the missing background fluctuations.  However, additional studies will be required to determine its true nature.

\section{acknowledgements}
We thank Alexander Kashlinsky for pointing out an error in the previous version of this manuscript.  We are grateful for our fruitful conversations with Michael Tremmel, Nir Mandelker, and Vivienne Baldasarre.  AR acknowledges funding from the NASA Earth and Space Science Fellowship (NESSF) grant number 80NSSC17K0459. FP acknowledges funding from the NASA Chandra award No. AR8-19021A and from the Nederlandse Onderzoekschool Voor Astronomie (NOVA).  NC and FP acknowledge NASA support through ADAP grant NNX16AF29G. 

\bibliography{ms}

\appendix

\section{Model Validation} \label{sec:model_validation}

Our model reproduces bolometric luminosity functions for $0<z<6$, as well as clustering of AGN to $z\sim 3$.  Recall that the set of merger trees used in the main text is calculated only for $z\geq 4$.  The following is based on a separate set of merger trees, which allow us to follow MBH growth to $z=0$ using parent halos ranging from $10^{10.6-15} \ \mathrm{M}_\odot$.  See \citet{Ricarte&Natarajan2018a,Ricarte&Natarajan2018b} for more details on this methodology.

\begin{figure*}
    \centering
    \includegraphics[width=\textwidth]{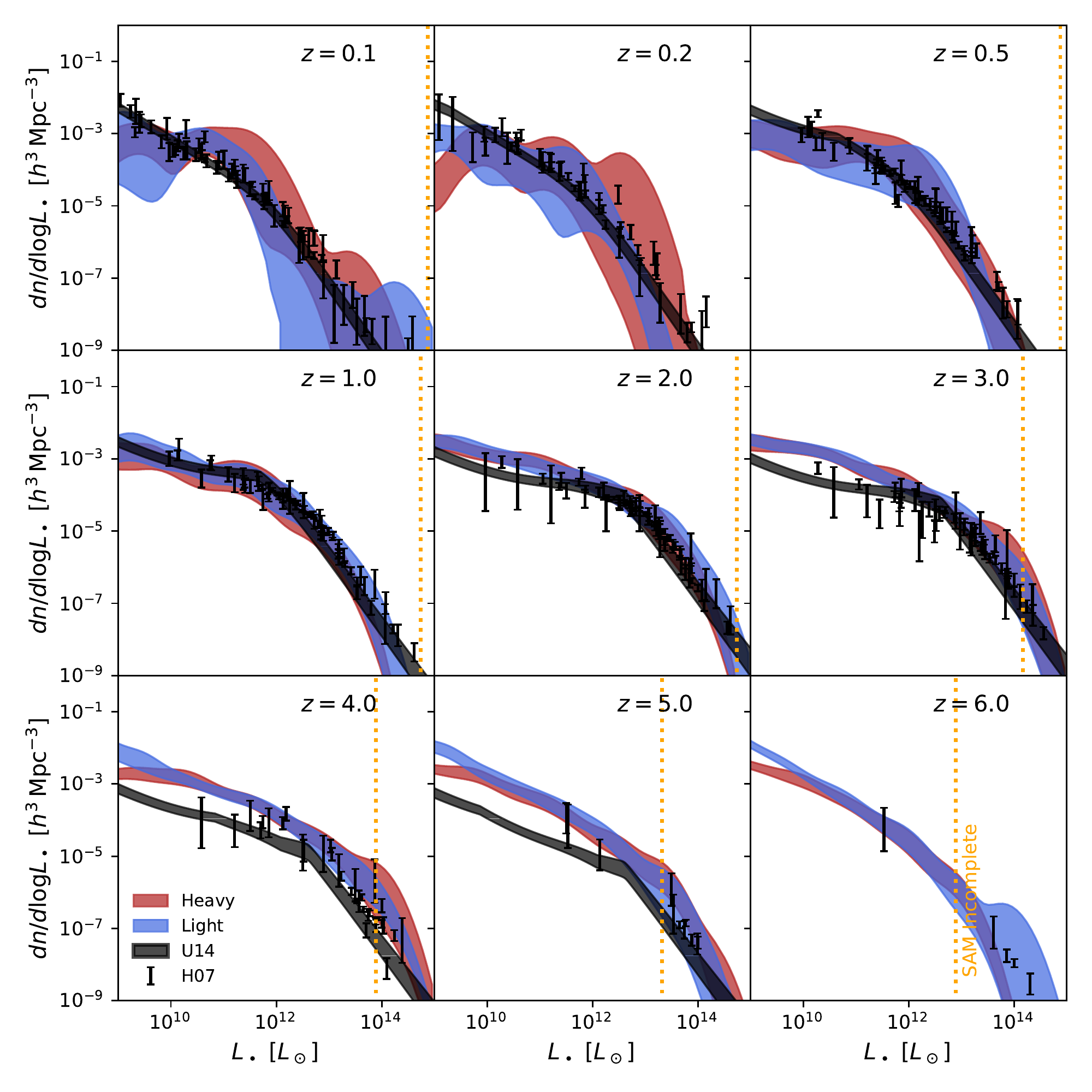}
    \caption{Bolometric luminosity functions from the model for both heavy and light seeds.  The orange dashed line indicates the luminosity beyond which half of haloes are expected to be missing from our dark matter merger trees.
    \label{fig:lf_backgrounds}}
\end{figure*}

Bolometric luminosity functions are displayed in Fig. \ref{fig:lf_backgrounds}, comparing to estimates from \citet{Hopkins+2007} and \citet{Ueda+2014} for the both the light and heavy seeds.  Our model agrees well for both seeds.  While our model overshoots somewhat at $z\geq 3
$, here the observations are highly uncertain, especially regarding the contribution of obscured BH growth to the faint end.  Since we are using a separate set of merger trees which go to $z=0$, rare massive peaks are missing at high-redshift.  An orange dashed line demarcates the luminosity beyond which we estimate half of haloes are missing from the dark matter merger trees.  At low redshift, artifacts appear and errors increase due to the decreased number of halos produced by the merger tree algorithm.

\begin{figure*}
    \centering
    \includegraphics[width=\textwidth]{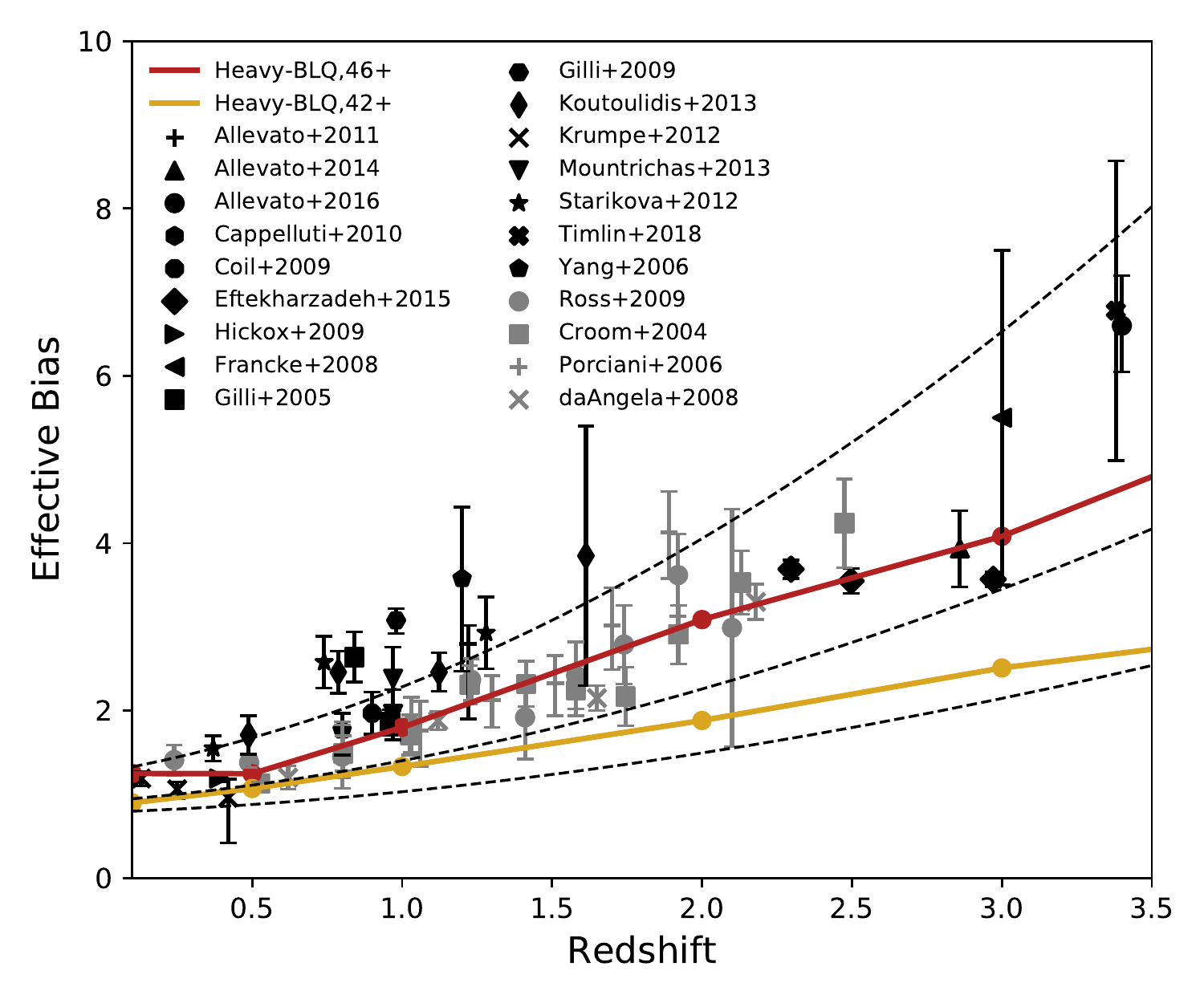}
    \caption{A compilation of X-ray AGN (black) and quasar (grey) clustering measurements compared to the effective biases computed by our model.  A fixed bolometric luminosity threshold, either $10^{46}$ (red) or $10^{42}$ (orange) $\mathrm{erg} \; \mathrm{s}^{-1}$, is selected.  Dashed lines correspond to the biases of haloes of $10^{11}$, $10^{12}$, and $10^{13} \ \mathrm{M}_\odot$.
    \label{fig:biases_many}}
\end{figure*}

AGN clustering is shown in Fig. \ref{fig:biases_many}, comparing to a wealth of observations for both X-ray selected AGN and broad-line quasars \citep{Timlin+2018,Allevato+2016,Eftekharzadeh+2015,Allevato+2014,Mountrichas+2013,Koutoulidis+2013,Starikova+2012,Krumpe+2012,Allevato+2011,Cappelluti+2010,Coil+2009,Ross+2009,Hickox+2009,Gilli+2009,daAngela+2008,Francke+2008,Porciani+2006,Yang+2006,Gilli+2005,Croom+2004}.  In red and orange, we show the clustering predicted by our model for AGN shining above $10^{46}$ and $10^{42}$ $\mathrm{erg} \; \mathrm{s}^{-1}$ bolometric luminosity thresholds.  These curves are computed following a traditional HOD modeling technique whereby the {\it number} of sources above these luminosity cuts are counted.  While an apples-to-apples comparison accounting for all of the selection effects that may affect AGN clustering measurements is beyond the scope of this paper, it is encouraging that our model fits the clustering of broad-line quasars well, as long as only luminous AGN above $10^{46} \ \mathrm{erg} \; \mathrm{s}^{-1}$ are included.  As the threshold is lowered, the clustering strength decreases, which is a natural prediction from any model in which AGN luminosity is correlated with halo mass.  While only heavy seeds are shown, the clustering of light seeds is similar, since these high-luminosity, high-mass AGN are insensitive to the seeding mechanism.

\section{Best Fits and Error Estimation}\label{sec:fitting}

Best fits correspond to the values obtained via $\chi^2$ minimization to a joint fit of infrared, X-ray, and cross-power spectra.  Since we lack a covariance matrix, all data points are treated independently.  This should not affect the best fit values, but may have an effect on our estimated errors.  We note that these fits are most strongly affected by the small error bars in the CIB power spectra on small scales.  At the same time, this region of the power spectrum is expected to have artifacts due to masking effects.  Therefore, we fit only on scales above 30 arcseconds and comment that including data on smaller scales leads to worse values of $\chi^2$ and a steeper slope $(\alpha=1.8)$ for the simple power law. 

We estimate errors using the emcee MCMC package \citep{Foreman-Mackey+2013}.  Uninformative priors are used for each of the parameters, ensuring only that the coherence does not exceed unity, for example.  Convergence is confirmed by visually inspecting both the evolution of both the median and standard deviation values of each chain's walkers.  Error regions are obtained by encapsulating 68 per cent of the posterior distribution and comparing to the best fit from the $\chi^2$ minimization.

\section{The Black Hole Mass Density}\label{sec:soltan}

Since black holes only radiate while they are accreting gas, any proposed flux attributed to AGN implies active black hole growth.  A hypothetical population of missing AGN implies that the overall black hole mass density must increase by this new growth channel.  Here, we derive the increase in the black hole mass density that is produced by a hypothetical population of AGN at a fixed redshift.

Suppose that the population emits a background in some band with a flux in units of $\mathrm{erg} \; \mathrm{s}^{-1} \; \mathrm{cm}^{-2} \; \mathrm{ster}^{-1}$

\begin{equation}
    \frac{dF}{d\Omega} = \frac{d^3 E}{dt dA d\Omega} ,
\end{equation}

\noindent and that at fixed redshift, the comoving (observed) flux per unit volume is 
\begin{equation}
    \frac{dF}{dV} = \frac{f_\mathrm{bol}}{4 \pi d_L^2} \frac{dL}{dV} ,
\end{equation}

\noindent where $f_\mathrm{bol}$ is the fraction of the bolometric luminosity radiated in the observed band. The observed flux can be expressed

\begin{equation}
    \frac{dF}{d\Omega} = \int \frac{dF}{dV}\frac{d^2V}{d\Omega dz}dz.
\end{equation}

Meanwhile, the Soltan argument \citep{Soltan1982} links the population of AGN to the mass density of black holes via

\begin{equation}
    \rho_\bullet = \frac{1-\epsilon}{\epsilon c^2} \int \frac{dL}{dV}\frac{dt}{dz}dz \, .
\end{equation}

Combining these results, the mass density can be linked to the observed flux and the population via

\begin{equation}
    \rho_\mathrm{extra} = \frac{1-\epsilon}{\epsilon}\frac{4\pi(1+z_\mathrm{eff})}{f_\mathrm{bol}c^3}\frac{dF}{d\Omega} \, ,
\end{equation}

\noindent where $z_\mathrm{eff}$ is the effective redshift of the population, defined as,

\begin{equation}
    1+z_\mathrm{eff} = \int (1+z) \frac{dF}{dV}\frac{d^2 V}{d\Omega dz} dz \bigg/ \int \frac{dF}{dV}\frac{d^2 V}{d\Omega dz} dz \, .
\end{equation}

From this equation, we can see that for any missing AGN population, the amount of extra black hole mass necessary to generate a fixed flux is substantial for all redshifts, scaling only as $(1+z)$.  AGN\_z2 and AGN\_z3 have $z_\mathrm{eff}$ of 2.37 and 3.10 respectively.  We assign our hypothetical populations of missing AGN $dF/d\Omega$ equal to the best fit in Table \ref{tab:background_parameters}, $f_\mathrm{bol}=0.1$, and $\epsilon=0.1$.  For both AGN\_z2 and AGN\_z3, the amount of extra black hole mass density gained during these AGN episodes is equal to

\begin{equation}
    \rho_\bullet = 10^{6.0} \ \mathrm{M}_\odot \mathrm{Mpc}^{-3},
\end{equation}

This is in excess of current estimates of the black hole mass density \citep{Shankar+2009}, and similar to that predicted by \citep{Yue+2013b}.

\end{document}